\documentclass[a4paper,11pt]{article}
\usepackage[utf8]{inputenc}

\usepackage{framed}
\usepackage{mdframed}

\usepackage{fullpage}
\usepackage[T1]{fontenc}
\usepackage{subfig}
\usepackage{tabularx}

\usepackage{epsf}
\usepackage{graphicx}
\usepackage{verbatim}
\usepackage{color}	
\usepackage{bbold}
\usepackage{hyperref}

\usepackage[british]{babel}
\usepackage{verbatim}
\usepackage[T1]{fontenc}
\usepackage{lmodern}
\usepackage{lipsum}
\usepackage{booktabs}
\usepackage{caption}
\usepackage{cite}
\usepackage{soul,color}
\usepackage[toc,page]{appendix}
\usepackage{mwe}

\usepackage{graphicx}
\usepackage{caption}

\usepackage{ifpdf}

\usepackage{float}

\usepackage{amsthm, amssymb}
\usepackage[tbtags]{amsmath}
\usepackage{bm}
\usepackage{mathtools}
\usepackage{amstext}
\usepackage{braket}
\usepackage{multirow}

\usepackage[normalem]{ulem}
\usepackage{xcolor}
\usepackage{cancel}

\newcommand\redsout{\bgroup\markoverwith{\textcolor{red}{\rule[0.5ex]{2pt}{0.4pt}}}\ULon}

\begin{document}

\vspace{5mm}
\vspace{0.5cm}

\def\be{\begin{eqnarray}}
\def\ee{\end{eqnarray}}

\def\ba{\begin{aligned}}
\def\ea{\end{aligned}}

\def\ls{\left[}
\def\rs{\right]}
\def\lc{\left\{}
\def\rc{\right\}}

\def\p{\partial}

\def\S{\Sigma}

\def\s{\sigma}

\def\O{\Omega}

\def\a{\alpha}
\def\b{\beta}
\def\g{\gamma}

\def\ad{{\dot \alpha}}
\def\bd{{\dot \beta}}
\def\gd{{\dot \gamma}}
\newcommand{\ft}[2]{{\textstyle\frac{#1}{#2}}}
\def\ib{{\overline \imath}}
\def\jb{{\overline \jmath}}
\def\Re{\mathop{\rm Re}\nolimits}
\def\Im{\mathop{\rm Im}\nolimits}
\def\trace{\mathop{\rm Tr}\nolimits}
\def\rmi{{ i}}

\def\N{\mathcal{N}}

\newcommand{\df}{h}
\newcommand{\dk}[1]{\frac{d^3#1}{(2\phi)^3}}
\newcommand{\Du}{\Delta_1}
\newcommand{\Dd}{\Delta_2}
\newcommand{\HH}{{\cal H}}
\newcommand{\UU}{{\cal U}}
\newcommand{\DD}{{\cal D}}
\newcommand{\A}{{\cal A}}
\newcommand{\Ad}{{\cal A}_\delta}
\newcommand{\At}{{\cal A}_\theta}
\newcommand{\Am}{{\cal A}_m}
\newcommand{\Bd}{{\cal B}_\delta}
\newcommand{\Bt}{{\cal B}_\theta}
\newcommand{\Bm}{{\cal B}_m}
\newcommand{\dd}{{\rm d}}
\renewcommand{\O}{{\cal O}}
\newcommand{\med}[1]{\langle #1\rangle}

\newcommand{\bea}{\begin{eqnarray}}
	\newcommand{\eea}{\end{eqnarray}}
\def\beq{\begin{equation}}
	\def\eeq{\end{equation}}
\newcommand{\hk}{\hat{k}}
\def\cK{{\cal{K}}}
\def\cI{{\cal{I}}}
\def\vp{\varphi}
\def\calo{{\cal O}}
\def\kq{{\kappa_q}}
\def\kqi{{\kappa_{q,i}}}
\def\kqj{{\kappa_{q,j}}}
\def\kk{{\kappa_k}}
\def\cals{{\cal S}}
\def\t{\eta}
\def\bx{\vec{x}}
\def\del{\partial}
\def\bnabla{\vec{\nabla}}
\def\d{{\rm d}}
\def\vW{{\vec{W}}}
\def\vk{{\vec{k}}}
\def\E{{\vec{E}}}
\def\vq{{\vec{q}}}
\newcommand{\ep}{\epsilon}
\newcommand{\GeV}{{\rm GeV}}
\newcommand{\TeV}{{\rm TeV}}
\newcommand{\eV}{{\rm eV}}
\newcommand{\half}{{\textstyle{\frac12}}}
\newcommand{\de}{\partial}
\newcommand{\al}{\alpha}
\newcommand{\bt}{\beta}
\newcommand{\gam}{\gamma}
\newcommand{\Gam}{\Gamma}
\newcommand{\Del}{\Delta}
\newcommand{\eps}{\varepsilon}
\newcommand{\zt}{\zeta}
\renewcommand{\th}{\theta}
\newcommand{\lam}{\lambda}
\newcommand{\Lam}{\Lambda}
\newcommand{\sig}{\sigma}
\newcommand{\vphi}{\varphi}
\newcommand{\om}{\omega}
\newcommand{\Om}{\Omega}
\newcommand{\rmd}{\textrm{d}}
\newcommand{\nab}{\nabla}
\newcommand{\Mb}{\bar{M}}
\newcommand{\Mpl}{M_{\textrm{Pl}}}
\newcommand{\fnl}{f_{\textrm{NL}}}
\newcommand{\avg}[1]{\Big< #1 \Big>}
\renewcommand{\k}{\vec{k}}
\newcommand{\q}{{\vec{q}}}
\newcommand{\R}{\mathcal{R}}
\newcommand{\pp}{p}
\newcommand{\D}{\Delta}
\newcommand{\yy}{y}
\newcommand{\hh}{D_{v}}
\newcommand{\no}{\nonumber}
\newcommand{\mad}{\mathrm{d}}
\newcommand{\ML}{\mathcal{L}}
\newcommand{\prt}{\partial}
\newcommand{\dg}{\delta_g}
\newcommand{\dgs}{{\delta_{g,s}}}
\newcommand{\dA}{\delta^{(A)}}
\newcommand{\dB}{\delta^{(B)}}
\newcommand{\tA}{\theta^{(A)}}
\newcommand{\tB}{\theta^{(B)}}
\newcommand{\Q}{\mathcal Q}
\newcommand{\tnl}{ \tau_{NL} }
\newcommand{\Quu}{Q_{11}}
\newcommand{\Qdd}{Q_{22}}
\newcommand{\Qud}{Q_{12}}
\newcommand{\Qut}{Q_{13}}
\newcommand{\Qdt}{Q_{23}}
\newcommand{\Qtt}{Q_{33}}
\newcommand{\expect}[1]{\left\Big< #1 \right\Big>}
\def\e{\zeta}
\def\Bs{{\overline{B}}_s}
\newcommand{\sla}{\slash \hspace{-0.2cm}}
\newcommand{\slam}{\slash \hspace{-0.25cm}}
\def\beqa{\begin{eqnarray}}
	\def\wm{\widetilde{m}}
	\def\arh{a_{\rm RH}}
	\def\am{a_{\rm m}}
	\def\eeqa{\end{eqnarray}}
\def\lsim{\mathrel{\rlap{\lower4pt\hbox{\hskip0.5pt$\sim$}}
		\raise1pt\hbox{$<$}}}         
\def\gsim{\mathrel{\rlap{\lower4pt\hbox{\hskip0.5pt$\sim$}}
		\raise1pt\hbox{$>$}}}         
\def\e{\zeta}
\def\cK{{\cal{K}}}
\def\cI{{\cal{I}}}
\def\vp{\varphi}
\def\calo{{\cal O}}
\def\kq{{\kappa_q}}
\def\kqi{{\kappa_{q,i}}}
\def\kqj{{\kappa_{q,j}}}
\def\kk{{\kappa_k}}
\def\cals{{\cal S}}
\def\t{\eta}
\def\bx{\vec{x}}
\def\del{\partial}
\def\bnabla{\vec{\nabla}}
\def\d{{\rm d}}
\def\vx{{\vec{x}}}
\def\vy{{\vec{y}}}
\def\vz{{\vec{z}}}
\def\vW{{\vec{W}}}
\def\vk{{\vec{k}}}
\def\E{{\vec{E}}}
\def\vq{{\vec{q}}}
\def\Bs{{\overline{B}}_s}
\def\wm{\widetilde{m}}
\def\pk{{\text{\tiny pk}}}
\def\arh{a_{\rm RH}}
\def\pbh{{\rm PBH}}
\def\am{a_{\rm m}}
\def\vp{{\vec{p}}}
\def\e{\delta}
\def\d{{\rm d}}
\def\lisa{\text{\tiny LISA}}
\def\PBH{\text{\tiny PBH}}
\def\M{{\tiny M}}
\newcommand{\F}[1]{\widehat{#1}}
\newcommand{\FPsi}{\F{\Psi}}
\newcommand{\FS}{\F{\mathcal S}}
\newcommand{\Ic}{\mathcal{I}_c}
\newcommand{\Is}{\mathcal{I}_s}
\newcommand{\Ics}{\mathcal{I}_{c,s}}
\newcommand{\T}{\mathcal T}
\newcommand{\Pz}{\mathcal P_\zeta}
\newcommand{\Ph}{\mathcal P_h}
\newcommand{\Bh}{S_h}
\newcommand{\vdelta}{\delta^{(3)}}
\renewcommand{\S}{\mathscr S}
\newcommand{\OGW}{\Omega_\text{GW}}
\newcommand{\rhoGW}{\rho_\text{GW}}
\newcommand{\etaf}{\eta_f}
\newcommand{\trA}{0.8}
\newcommand{\trB}{0.65}
\renewcommand{\dh}{\delta h}
\newcommand{\hc}{h_\textrm{c}}
\newcommand{\te}{t_\textrm{e}}
\newcommand{\colboxed}[1]{#1}
\def\eeqa{\end{eqnarray}}


\def\la{~\mbox{\raisebox{-.6ex}{$\stackrel{<}{\sim}$}}~}
\def\ga{~\mbox{\raisebox{-.6ex}{$\stackrel{>}{\sim}$}}~}
\def\bq{\begin{quote}}
\def\eq{\end{quote}}
\def\PL{{ \it Phys. Lett.} }
\def\PRL{{\it Phys. Rev. Lett.} }
\def\NP{{\it Nucl. Phys.} }
\def\PR{{\it Phys. Rev.} }
\def\MPL{{\it Mod. Phys. Lett.} }
\def\IJMP{{\it Int. J. Mod .Phys.} }
\font\tinynk=cmr6 at 10truept
\newcommand{\arXiv}[2]{\href{http://arxiv.org/pdf/#1}{{\tt [#2/#1]}}}
\newcommand{\arXivold}[1]{\href{http://arxiv.org/pdf/#1}{{\tt [#1]}}}

\newcommand{\llp}{\left [}
\newcommand{\rrp}{\right ]}

\newcommand{\lp}{\left (}
\newcommand{\rp}{\right )}

\newcommand{\ik}{\int \frac{\d ^3 k}{(2 \pi )^{3}}}
\newcommand{\eikx}{e^{-i \vk \cdot \vx}}

\newcommand{\ikp}{\int \frac{\d ^3 k'}{(2 \pi )^{3}}}

\newcommand{\ipo}{\int \frac{\d ^3 p_1}{(2 \pi )^{3}}}
\newcommand{\epo}{e^{-i \vec p_1 \cdot \vx}}
\newcommand{\vpo}{\vec p_1 }

\newcommand{\ipt}{\int \frac{\d ^3 p_2}{(2 \pi )^{3}}}
\newcommand{\vpt}{\vec p_2 }

\newcommand{\ipr}{\int \frac{\d ^3 p_3}{(2 \pi )^{3}}}
\newcommand{\vpr}{\vec p_3 }  

\newcommand{\ipf}{\int \frac{\d ^3 p_4}{(2 \pi )^{3}}}
\newcommand{\vpf}{\vec p_4 }  

\newcommand{\ipc}{\int \frac{\d ^3 p_5}{(2 \pi )^{3}}}
\newcommand{\vpc}{\vec p_5 }  

\newcommand{\dek}{\delta_k}
\newcommand{\Rk}{\R_k}
\newcommand{\Rkp}{\R_{k'}}

\renewcommand{\P}{\mathcal{P}}

\def\hc{c.c.}

\numberwithin{equation}{section}

\allowdisplaybreaks

\allowbreak


\begin{titlepage}

\hfill\text{CERN-TH-2023-078}

\vspace{0.005cm}

\hfill\text{LAPTH-018/23}

\begin{flushright}


\end{flushright}

\begin{center}
	    { \LARGE{
	    \textbf{The Weak Gravity Conjecture, Overcharged Shells   \\
     \vskip 0.2cm
      and  Gravitational   Traps} 
	    }}

		\vspace{40pt}

		{\Large Alex~Kehagias$^{a,b}$,\,
         Kostas\,D.~Kokkotas$^{c}$,\,
         Antonio~Riotto$^{d,e}$,\, \\ 
         \vskip 0.2cm
         John~Taskas$^{a}$ and George~Tringas$^{e}$}

		\vspace{30pt}

{

$^{a}$
{Physics Division, National Technical University of Athens,\\
15780 Zografou Campus, Athens, Greece}
\vspace{12pt}

$^{b}$
{CERN, Theoretical Physics Department, Geneva, Switzerland}
\vspace{12pt}

$^{c}$
{Theoretical Astrophysics, IAAT,\\
University of T\"ubingen, 72076 T\"ubingen, Germany}
\vspace{12pt}

$^{d}$
{Département de Physique Théorique and Gravitational Wave Science Center (GWSC),\\
Université de Genève, 24 quai E. Ansermet, CH-1211 Geneva, Switzerland}
\vspace{12pt}

$^{e}$
{Laboratoire d'Annecy-le-Vieux de Physique Théorique {(LAPTh)},\\
CNRS, Université Savoie Mont Blanc (USMB), UMR 5108, \\
9 Chemin de Bellevue, 74940 Annecy, France }

}

\vspace{30pt}

{ABSTRACT}

\end{center}

\vspace{1pt}
\noindent
The Weak Gravity Conjecture predicts that in quantum gravity there should exist overcharged states, that is states with charge larger than their mass. Extending this to large masses and charges, we are expecting similar overcharged classical solutions. This has been demonstrated in higher-derivative extensions of General Relativity. In this paper we investigate  the existence of overcharged solutions in General Relativity.  We study the dynamics  of a thin shell of mass $m$ and charge $Q$ under the action of its own gravitational and $U(1)$ fields. We show that  shells with surface energy $\sigma$ and pressure $P$ obeying $P=w\sigma$ with $0\leq w\leq 1$ are necessarily undercharged $m\geq |Q|$ and  always collapse to form  Reissner-Nordstr\"om black holes. Nevertheless, if $-1\leq w<0$, we find that overcharged  $m\leq |Q|$  shells exist, which however,   are inevitably  stabilized at finite radial distance.
Therefore they never form  naked singularities in accordance with cosmic censorship and the conjectured relation between cosmic censorship and the Weak Gravity Conjecture. 
An intriguing consequence of the existence of such overcharged shells is that  gravitational modes may form bound states   due to the peculiar form of the Regge–Wheeler–Zerilli potential. This might  lead  to   gravitational traps close  the  surface of near-overcharged shells.



\vspace*{\fill}
\noindent
{\rm \footnotesize E-mails: kehagias@central.ntua.gr\,,\,kostas.kokkotas@uni-tuebingen.de\,,\,antonio.riotto@unige.ch\,,\\
taskas@mail.ntua.gr\,,\,tringas@lapth.cnrs.fr}

\end{titlepage}


{\hypersetup{hidelinks}
\tableofcontents
}

\setcounter{footnote}{0}

\baselineskip 5.6 mm



\section{Introduction}

Although not much are known for theories of quantum gravity, there are some  remarkably simple statement  about principles that such theories should respect \cite{V,P,Har}.  The Weak Gravity Conjecture (WGC) is one of these principles \cite{AHMNV}. It argues that any gauge force must be stronger than gravity. Besides its simplicity, this statement has profound consequences which have been extensively  been explored.   
One of them is that any consistent quantum theory of gravity should contain in its spectrum states of mass $m$ and charge $Q$ that satisfy $m< |Q|$ (in appropriate units). Elementary particles are examples of such states.

The above argument   extends also  to states with mass larger than the Planck scale, which are believed to be described as  black holes (BHs). 
In particular, 
the mass-to-charge ratio   $\alpha=m/|Q|=1$ obeyed by extremal BHs cannot be exact and should decrease with $Q$ so that for any extremal BH there is another one with $\alpha<1$. In the opposite case, the extremal BHs would be absolutely stable. Extending the above argument to large values of $m$ and $Q$ that describe macroscopic BHs, it is expected that there are spherical symmetric, charged  objects that obey $m/|Q|<1$. Such  objects have been shown to exist in higher-derivative extensions of GR. The latter are supposed to encode the corrections due to quantum gravity effects and presumably will allow for BH solutions with $m<|Q|$.  Indeed, it has been shown in \cite{motl} that the order $\mathcal{O}(Q^{-4})$ effective action will have
the generic structure 
\begin{align}
   \mathcal{S}= \int d^4x
& \sqrt{-g} \bigg( \frac{R}{2\kappa^2} -
\frac{1}{4}F_{\mu\nu}F^{\mu\nu}
+ c_1\,R^2 + c_2\, R_{\mu\nu}R^{\mu\nu} + c_3\,R_{\mu\nu\rho\sigma}R^{\mu\nu\rho\sigma} +  \nonumber \\
&  + \, c_4\,RF_{\mu\nu}F^{\mu\nu} +
c_5\,R^{\mu\nu}F_{\mu\rho}{F_\nu}^\rho +
c_6\,R^{\mu\nu\rho\sigma}F_{\mu\nu}F_{\rho\sigma} +
c_7\,(F_{\mu\nu}F^{\mu\nu})^2  \nonumber \\
&  + \, c_8\,(\nabla_{\mu} F_{\rho\sigma})(\nabla^{\mu}
F^{\rho\sigma}) + c_9\,(\nabla_{\mu} F_{\rho\sigma})(\nabla^{\rho}
F^{\mu\sigma}) \bigg)\, . 
\end{align}
The masses of the extremal BH solutions of the standard two-derivative GR are corrected such that $m<|Q|$, supporting the WGC. However, the question remains: are there solutions in GR that allow for $m<|Q|?$ 

The goal of this paper is to show that indeed in GR there are extended spherical symmetric objects that have charge larger that their mass. We consider spherical symmetric shells within GR with \textit{total energy} $m$ and charge $Q$  and appropriate Israel matching conditions.   The  surface energy-momentum stress tensor is that of a perfect fluid with equation of state $P=w\sigma$, where $P,\sigma$  are the surface pressure and tension, respectively.    We find that when $0\leq w\leq 1$,   necessarily $m\geq |Q|$. In this case, the shell  always collapses under its own gravitational field forming a Reissner-Nordstr\"om  black hole. However, for $-1\leq w<0$, there are overcharged $m<|Q|$ shells, which however are  stabilized at some finite radius. This is  consistent with the cosmic censorship hypothesis, revealing the connections between the latter and the WGC, which has been advocated in \cite{HS1,HS2}. The cosmic censorship for overcharged shells has also been discussed in \cite{Huben}. Overcharged brane shells have also been exploited in 
\cite{vR1} from compactifications of type IIB string theory, whereas,  the stability of spherical shells of matter in Newtonian gravity and general relativity 
has been discussed in \cite{kuchar,Chase,LE1,SF1,SF2,PER1,FL1}.  Thereby, extended stable objects satisfying the WGC and consistent with the cosmic censorship exist already in  classical Einstein-Maxwell theory.


\section{Charged shells in Newtonian gravity} \label{newt}
Let us see what we can learn, based on Newtonian dynamics, for  the gravitational  collapse of a thin spherical shell with charge $Q$ and rest mass $M$. Its energy $m$ is the sum of the  rest mass, the Coulomb and the gravitational binding energy so that 
\begin{eqnarray}
m=M+\frac{Q^2}{r}-\frac{m^2}{r}\,. \label{mm0}
\end{eqnarray} 
We see that the energy is a  function of the distance $r$ from the center and its derivative is given by
\begin{eqnarray}
m'(r)=\frac{m^2-Q^2}{r(2m+r)}\,. \label{md}
\end{eqnarray}
The shell will collapse if  $m$ decreases with
decreasing $r$. Therefore, from (\ref{md}), the shell collapses when
it is undercharged $m^2>Q^2$, whereas it expands forever in the  opposite overcharged case $m^2<Q^2$. 

In the above argument, we have assumed that the internal energy $U$ of the shell is negligible. However, we should also take into account $U$ in a full treatment of the energy budget of the problem. Hence, the total energy of the shell should be 
\begin{eqnarray}\label{totalenergy1}
m&=&M+\frac{Q^2}{r}-\frac{m^2}{r}+U\, . \label{mm00}
\end{eqnarray}
In the simplest case,  we may  consider a shell supporting an adiabatic fluid on its surface with surface energy density $\sigma$, pressure $P$ and equation of state $P=w\sigma$. In this case, the internal energy satisfies
\begin{eqnarray}
\text{d}U=-P\,\text{d}A \label{U}\,,
\end{eqnarray}
where $\text{d}A=8\pi r \text{d}r$ is the surface element on the spherical shell. Eq.(\ref{U}) represents the fact that the increase of the surface energy compensates the rate of work done by the pressure for expanding the shell. For adiabatic process, we have that 
\begin{eqnarray}
P(4\pi r^2)^\gamma=P_0(4\pi r_0^2)^\gamma\,, 
\end{eqnarray}
so that 
\begin{eqnarray}\label{adiabaticP}
 P=P_0\left(\frac{r_0}{r}\right)^{2\gamma}\,,  
\end{eqnarray}
where $\gamma=w-1$ and $r_0,~P_0$ are the radius and the pressure of some initial shell configuration, respectively. \footnote{Similarly, since $P=w\sigma$ we get that 
$\sigma=\sigma_0 \left(\frac{r_0}{r}\right)^{2\gamma}$, 
where $\sigma_0=P_0/w$. In addition, integrating Eq.(\ref{U}) and using Eq.\eqref{adiabaticP} we find that
$U=\frac{4\pi}{w-2}\, P \,r^2
\,$ and $U_0=4\pi \sigma_0r_0^{2\gamma}\left(\frac{r_0}{r}\right)^{2w}\,.$} Therefore, the energy $m$ when the internal energy $U$ is taken into account according to Eq.(\ref{mm00}), turns out to be
\begin{align}
m'(r)&=\frac{m^2-Q^2+r^2\,U'(r)}{r(2m+r)}\nonumber \\
&=\frac{m^2-Q^2-8\pi\,P\,r^3}{r(2m+r)}\,.
\end{align}
The condition then $m'>0$ for an imploding shell 
is equivalent to 
\begin{eqnarray}
m^2-Q^2> 8\pi \, P\,  r^3. \label{mqu}
\end{eqnarray} 
We see from Eq.(\ref{mqu}) that for $P\geq0$ (i.e., $w\geq0$) we have $m^2-Q^2\geq 0$. In other words:  Collapsing shells with $w\geq 0$ are necessarily undercharged $m\geq |Q|$.
 
However, for $w<0$ things are different: In this case, the pressure is negative $p<0$ and clearly we may have imploding shells with $m^2<Q^2$. Therefore, we have collapse only when
\begin{eqnarray}
&&m\geq|Q|~~~~\mbox{and }~~~~w\geq 0\,,  \nonumber \\
&&m<|Q|~~~~\mbox{and }~~~~w< 0\,. \label{bounds}
\end{eqnarray}
Note that for $w<0$, there is also maximum value of $|Q|$ since from 
Eq.(\ref{mqu}) we get 
\begin{eqnarray}
|Q|<\sqrt{m^2+8\pi|P_0|r_0^3}\,. 
\end{eqnarray}
What we learned so far is that gravitational collapse occurs for undercharged, $m\geq |Q|$ as well as for overcharged,  $m<|Q|$ shells provided that in the latter case the shell is composed of adiabatic fluid with negative pressure. The collapse of course will eventually end up in the formation of a Reissner-Nordstr\"om black hole for an undercharged shell, and the formation of a naked singularity violating cosmic censorship for an overcharged one. Below we will examine the latter possibility, namely the collapse of an overcharged $|Q|>m$ shell, and look for its ultimate  fate. In particular, we want to see if such a shell inevitable  form a naked singularity  violating cosmic censorship.  

\section{Charged shells in General Relativity}

Let us now examine the dynamics of the charged shell within General Relativity.  To be more specific, let us consider a charged spherical thin shell of total rest mass $M$, total charge $Q$, energy density $\sigma$ and pressure $p$. The intrinsic metric on the shell is
\begin{eqnarray}
\d s_3^2=\gamma_{ij}\d \xi^i\d \xi^j=-\text{d}\tau^2+R^2\d \Omega^2,~~~(i,j=0,1,2) \,, \label{mm}
\end{eqnarray}
where $\xi^i=(\tau,\theta,\phi)$ are the shell worldvolume coordinates. The scale factor $R=R(\tau)$ is the radius of the shell and is a function of the proper time $\tau$ where the later is the time measured by an observed moving along with the shell. 
The stress-energy tensor  $S_{ij}$ of the shell has the form of a perfect fluid 
\begin{eqnarray}
    S_{ij}=(\sigma+p)u_iu_j+P\gamma_{ij},
\end{eqnarray}
where $u^i$ is the three-velocity of the shell constituents.
Its non-zero components are explicitly written as 
has components
\begin{align}
   S^{0}_{0}=-\left(\frac{1}{2}\sigma+P\right), ~~~~
   S^1_1=S^2_2=\frac{1}{2}\sigma.
\end{align}
Then, energy-momentum conservation leads to the equation \cite{Chase}
\begin{eqnarray}
\nabla_\alpha (\sigma u^\alpha)=-P \nabla_\alpha u^\alpha \,, \label{ce}
\end{eqnarray}
Eq.(\ref{ce}) can be written as 
\begin{eqnarray}
P+\frac{\sigma}{2}+\frac{1}{2}\frac{\text{d}(\sigma R)}{\text{d}R}=0 \,,  \label{ce1}
\end{eqnarray}
Note that Eq.(\ref{ce1}) is nothing else than 
\begin{eqnarray}
\text{d}M=-P\,\text{d}A=-8\pi P R\,\text{d}R\, ,  \label{dm}
\end{eqnarray}
where 
\begin{eqnarray}
    M=4\pi R^2 \sigma
\end{eqnarray}
is the energy of the shell. 
For a perfect fluid  with an equation of state $p=w \sigma$, by solving Eq.(\ref{ce1}) we find the general relations
\begin{eqnarray}
\sigma=\frac{\sigma_0}{R^{2(1+w)}}\,,
~~~~~P=\frac{P_0}{R^{2(1+w)}}\,,~~~~ M=\frac{M_0}{R^{2w}}\,, 
~~~~M_0=4\pi \sigma_0\,, ~~~~~P_0=w\sigma_0.\label{mass}
\end{eqnarray}
The spacetime in the exterior of the shell is described by the  Reissner-Nordstr\"om geometry with line element
\begin{eqnarray}
\d s_+^2=g_{\mu\nu}^+\d x^{\mu}_+\d x^{\nu}_+=-f(r) \d t^2+\frac{\d r^2}{f(r)}+r^2 \d\Omega_+^2\,, ~~~~~f(r)=1-
\frac{2m}{r}+\frac{Q^2}{r^2}\,, \label{met+}
\end{eqnarray}
and Maxwell field strength, 
\begin{eqnarray}
F_{rt}=\frac{Q}{r^2}\,,
\end{eqnarray}
where $m$ and $Q$ are the total energy (gravitational mass) and charge, respectively. For completeness we write down the coordinates of the external manifold $x^{\mu}_+=(t,r,\theta,\phi)$ and the inner $r_-$ and outer $r_+$ horizons at radius
\begin{align}\label{horizons}
    r_{\pm}=m\pm\sqrt{m^2-Q^2} \,.
\end{align}
Inside the shell, the \textit{internal} spacetime is Minskowski with vanishing electromagnetic field
\begin{eqnarray}
\d s^2_-=g_{\mu\nu}^-\d x^{\mu}_-\d x^{\nu}_-=-\d \eta^2+\d \chi^2+\chi^2 \d \Omega_-^2\,, \label{met-}
\end{eqnarray}
where $\eta$ and $\chi$ are time and distance coordinate in the interior of the shell and $x^{\mu}_-=(\eta,\chi,\theta,\phi)$.

The induced metrics $\gamma^{\pm}_{ij}$ on the shell are calculated by the pullback of the interior and exterior metric $g_{\mu\nu}^{\pm}$ on the hypersurface
\begin{align}
    g_{\mu\nu}^{\pm}(e_{\pm})^{\mu}_{\,\,i}(e_{\pm})^{\nu}_{\,\,j}=\gamma^{\pm}_{ij} \,,\quad\quad (e_{\pm})^{\mu}_{\,\,i}=\frac{\partial x_{\pm}^{\mu}}{\partial\xi^i} \,,
\end{align}
with $(e_{\pm})^{\mu}_{\,\,i}$ the tangent vector to the hypersurface. The line element from the pullback of the \textit{interior} side is
\begin{align}
    \d\sigma^2_-=-\left(\dot{t}^2_--\dot{R}^2_-\right)\d\tau^2+R^2_-\d\Omega^2_2 \,, \label{mmi}
\end{align}
while from the \textit{exterior}
\begin{align}
    \d\sigma^2_+=-\left(f(R_+)\dot{t}^2_+-\frac{\dot{R}^2_+}{f(R_+)}\right)\d\tau^2+R^2_+\d\Omega^2_2 \,.\label{mme}
\end{align}
The dot stands for derivative with respect to the shell proper time $\tau$.
The metrics in Eqs. (\ref{mmi}) and (\ref{mme}) should coincide with the shell metric Eq. (\ref{mm})  
so that
\begin{eqnarray}
\dot{t}^2_--\dot{R}^2&=&1 \quad\mathrm{(interior)}\,,\label{Kostas1}\\
~f(R)\dot{t}^2_+-\frac{\dot{R}^2}{f(R)}&=&1 \quad \mathrm{(exterior)}\,,\label{Kostas2}\\
R_- = R_+ &=& R \,.\label{Kostas3}
\end{eqnarray}
Next we introduce the three-dimensional tensor $K_{ij}^{\pm}$ which stands for the extrinsic curvature defined by the pullback of the change of the normal vector $n^{\mu}$ along each side of the hypersurface
\begin{align}
    K_{ij}^{\pm}=(e_{\pm})^{\mu}_{\,\,i}(e_{\pm})^{\nu}_{\,\,j}\nabla_{\mu}n_{\nu}^{\pm}\,. 
\end{align}
The extrinsic curvature is not continues but it has jumps specified by the  the Israel matching condition which is explictly  written as \cite{CI}
\begin{align}\label{Condition2}
    \Delta K_{ij}=K^+_{ij}-K^-_{ij}=-8\pi\left(S_{ij}-\frac{1}{2}S^{\,k}_k\gamma_{ij}\right)\,,
\end{align}


Next consider the four-velocity vector $U^{\mu}$ and the normal vector $n^\mu$ to the shell surface defined as 
\begin{align}
    u^{\mu}=\frac{\d x^{\mu}}{\d\tau}\,,\,\,\,\,\,\,\,\,\,\,\,\,
    u^{\mu}u^{\nu}g_{\mu\nu}=-1\,,\,\,\,\,\,\,\,\,\,\,\,\,
    n^{\mu}u_{\mu}=0\,,\,\,\,\,\,\,\,\,\,\,\,\,
    n^{\mu}n_{\mu}=1\,,\,\,\,\,\,\,\,\,\,\,\,\,
    n_{\mu}e^{\mu}_{\,\,i}=0\,.
\end{align}
Solving Eqs.\eqref{Kostas1}-\eqref{Kostas2} for $\dot{t}_{\pm}$, the four-velocities are calculated to be
\begin{align}
    u^{\mu}_-=\left(\sqrt{1+\dot{R}^2},\dot{R},0,0\right)\,,\quad\quad u^{\mu}_+=\left(f^{-1}\sqrt{f+\dot{R}^2},\dot{R},0,0\right)\,.
    \label{uu}
\end{align}
whereas the normal vector turns out to be 
\begin{align}
    n_{-}^{\mu}=\left(\dot{R},\sqrt{1+\dot{R}^2},0,0\right)\,,\quad\quad
    n_{+}^{\mu}=\left(\dot{R},f^{-1}\sqrt{f+\dot{R}^2},0,0\right)\,.
    \label{nn}
\end{align}
Then,  the Israel matching condition (\ref{Condition2}) is written as 
\begin{eqnarray}
    n_\mu^+\frac{du^\mu_+}{d\tau}- n_\mu^-\frac{du^\mu_-}{d\tau}
    =8\pi \left(\frac{1}{2}\sigma+p\right).  \label{un}
\end{eqnarray}
Using Eqs. (\ref{uu}) and (\ref{nn}), the Israel matching condition 
in Eq. (\ref{un}) is explicitly written as 
\begin{eqnarray}
\Big(f+\dot R^2\Big)^{1/2}-\Big(1+\dot R^2\Big)^{1/2}=-4\pi \sigma R\,.  \label{israel}
\end{eqnarray}
By using the explicit form of $f$, Eq. (\ref{israel}) specifies the total energy $m$ to be
\begin{eqnarray}
m=M\Big(1+\dot{R}^2\Big)^{1/2}-\frac{M^2-Q^2}{2R}\,. \label{mm1}
\end{eqnarray}
which remains constant during the evolution with proper time $\tau$ while $M$ and $R$ change. Therefore, $m$ can be determined at the equilibrium point (if such a point exists) and remain constant through the whole dynamics. 

We will now search for stable equilibrium points, i.e. points where the shell can be stabilized. 
The conditions for stable equilibrium at $R=R_0=const.$  is
\begin{eqnarray}
\dot{R}\big|_{R_0}=0, ~~~~~\ddot{R}\big|_{R_0}>0\,,
\end{eqnarray}
which leads to, with $m_0=m(R_0)$, 
\begin{eqnarray}
m_0&=&M-\frac{M^2-Q^2}{2R_0}\,,\label{eq1}\\
\frac{\partial m_0}{\partial R_0}
&=&\frac{d M}{d R_0}\left(1-\frac{M}{R_0}\right)
+\frac{M^2-Q^2}{2R_0^2}=0\,, \label{eq2}\\
\frac{\partial^2 m_0}{\partial R_0^2} &>&0\,. \label{eq3}
\end{eqnarray}
The conditions above ensure the stability of the shell at specific radius $R_0$. At this radius the shell is gravitationally stable. Note that, if such a stable point $R_0$ exists,  the shell is always stabilized outside the outer horizon of the exterior metric, i.e.
\begin{eqnarray}\label{fR}
f(R_0)=1-\frac{2m}{R_0}+\frac{Q^2}{R_0^2}=\left(1-\frac{M}{R_0}\right)^2>0\,.\label{eq4}
\end{eqnarray} 
 Once the outer horizon is crossed, the fate of the shell is predetermined: it will end up in a charged BH.

\subsection{Charged shells with zero or positive pressure}\label{subsection31}
To specify the equilibrium radius $R_0$ we proceed as follows. Once an equation of state $P=P(\sigma)$ is given, we can solve Eq.(\ref{dm}), in order to determine the mass $M=M(R)$. We may then use $M(R)$ in Eqs.(\ref{eq1}) and (\ref{eq2}) to find $R_0$. Finally, we check if this solution satisfies Eq.(\ref{eq3}). If it does, then the shell is at stable equilibrium at $R_0$.

In order to establish the above procedure, let us consider the known  case of a charged dust with $p=0$ \cite{kuchar}. In this case, we have from Eq.(\ref{mass})  
\begin{eqnarray}
\sigma=\frac{\sigma_0}{R^2} \,,
\end{eqnarray}
where $\sigma_0$ is a constant, so that 
\begin{eqnarray}
M=4\pi\sigma_0 \,.
\end{eqnarray}
It is then easy to see from Eqs.(\ref{eq1})-(\ref{eq4}) that the condition for equilibrium is satisfied for extremal shells, i.e, shells that satisfy  $m=|Q|$. However, such  shells are at neutral equilibrium since $\partial^2 m/\partial R_0^2=0$ for any $R_0$. Therefore, if the shell is initial at rest at $R_0$, it will remain there in indifferent equilibrium if $P=0$. In fact it has been proven in \cite{kuchar} that when $p\geq 0$, the shell can never be overcharged, i.e.,
\begin{eqnarray}
m\geq |Q| ~~~ \mbox{for} ~~~ P\geq 0 \,. \label{bps}
\end{eqnarray}
Then, such a shell will inevitably collapse once enters the outer Reissner-Nordstr\"om horizon. This is in accordance with the conclusion 
we arrived in section (\ref{newt}) based purely on Newtonian arguments.

To establish Eq. (\ref{bps}), we note that 
\begin{eqnarray}\label{pressure}
P=\frac{1}{16\pi}\frac{M^2-Q^2}{R^2\Big(R(1+\dot{R}^2)^{1/2}-M\Big)} \,.
\label{pp}
\end{eqnarray}
In order the pressure $P$ to be always positive ($P>0$) the motion should happen with radius $R$ larger than a critical $R_c$ (when  $M^2>Q^2$)
\begin{eqnarray}
R>R_c \,, ~~~ R_c=\frac{M}{(1+\dot{R}^2)^{1/2}} \,. \label{bb}
\end{eqnarray}
Then expanding the inequality 
\begin{eqnarray}
\left(M(1+\dot{R}^2)^{1/2}-|Q|\right)^2\geq 0 \,,
\end{eqnarray}
we find that 
\begin{eqnarray}
|Q|\leq \frac{1}{2}M (1+\dot{R}^2)^{1/2}+\frac{1}{2}\frac{Q^2}{M (1+\dot{R}^2)^{1/2}} \,.
\label{m01}
\end{eqnarray}
In addition, from Eq.(\ref{mm1}) we see that $m=m(R)$ is an increasing function of $R$ and therefore
\begin{align}
m(R_c)\leq m \,.
\end{align}
On the other hand, for $R=R_c$ we have   
\begin{align}
m(R_c)=& \,M(1+\dot{R}^2)^{1/2}-\frac{M^2-Q^2}{2R_c}\nonumber \\
=&\,\frac{1}{2}M (1+\dot{R}^2)^{1/2}
+\frac{1}{2}\frac{Q^2}{M}(1+\dot{R}^2)^{1/2} \,.
\label{m002}
\end{align}
Adding and subtracting the second term of Eq.(\ref{m002}),  we can write the above equation as 
\begin{eqnarray}
m(R_c)= \frac{1}{2}\frac{Q^2}{M}(1+\dot{R}^2)^{1/2}-\frac{1}{2}\frac{Q^2}{M (1+\dot{R}^2)^{1/2}}+\left(\frac{1}{2}M(1+\dot{R}^2)^{1/2}+\frac{1}{2}\frac{Q^2}{M (1+\dot{R}^2)^{1/2}}
\right) \,.
\label{m02}
\end{eqnarray}
Therefore we get 
\begin{align}
m(R_c)&\geq |Q| 
+\frac{1}{2}\frac{Q^2}{M}(1+\dot{R}^2)^{1/2}
\left(1-\frac{1}{1+\dot R ^2}\right)\nonumber\\
&\geq |Q|+\frac{1}{2}\frac{Q^2}{M}\frac{\dot R^2}{(1+\dot{R}^2)^{1/2}}\geq |Q|
\label{0221}
\end{align}
and using Eq.(\ref{m02}) we get that bound 
\begin{eqnarray}
m\geq |Q| \,.
\end{eqnarray}
Therefore, there are \textit{no overcharged shells} for $P\geq 0$ ($w\geq0$). It is clear now that the above argument cannot work whenever the pressure $p$ in Eq.(\ref{pp}) is negative $P<0$  ($-1\leq w<0$),  in accordance with the general discussion based on purely Newtonian considerations. 

\subsection{Charged shells with negative pressure}\label{w14}

Using Eq. (\ref{mass}) for a general equation of state $p=w\sigma$, the total energy at equilibrium $R=R_0$ is
\begin{align}\label{mgeneral}
    m_0 =\frac{Q^2}{2R_0}-\frac{M_0}{2R^{1+4w}_0}\left(M_0-2R_0^{1+2w}\right)\,,
\end{align}
where the first term stands for the electric energy potential, the second for the surface tension contribution and the third for gravitational potential\footnote{For completeness we write down the general formula for the total energy in Eq.\eqref{mm1} in terms of $M_0$ using Eq.\eqref{mass}
\begin{align}\label{mt}
    m=M_0R^{-2w}\sqrt{1+\dot{R}^2}+\frac{Q^2-M_0^2R^{-4w}}{2R} \, ,
\end{align}
where is this expression $R\equiv R(t)$.}. Although we can keep the discussion general, we prefer to work out  a particular value for $w$ for which analytic results can be obtained. A particular simple case which can be treated completely analytically is  $w=-1/4,$ . In this case, using Eq.\eqref{mgeneral} we find that the critical point condition in Eq.\eqref{eq2} specifies the equilibrium point at $R=R_0$ 
\begin{eqnarray}
\frac{\partial m_0}{\partial R_0}=\frac{M^2_0R_0^{3/2}-Q^2}{2R_0^2}=0  \,,
\end{eqnarray}
so that the equilibrium point is at distance
\begin{eqnarray}\label{R0}
R_0=\left(\frac{Q^2}{M_0}\right)^{2/3},
\end{eqnarray}
from the center of the shell. In addition, this is a stable equilibrium point since 
\begin{eqnarray}
\frac{\partial^2m_0}{\partial R^2_0}=\frac{3M_0^2}{4Q^2}>0\,.
\end{eqnarray}
The total gravitational energy at the equilibrium point $R_0$ turns out to be
\begin{eqnarray}\label{m014}
m_0=-\frac{1}{2}M_0^2+\frac{3}{2}\Big(M_0 Q\Big)^{2/3}\,,
\end{eqnarray}
which can be written as 
\begin{eqnarray}
Y=-\frac{1}{2}Z^2+\frac{3}{2}Z^{2/3}\,,
\end{eqnarray}
where 
\begin{eqnarray}
Z=\frac{M_0}{|Q|^{1/2}} ~~~ \mbox{and} ~~~ Y=\frac{m_0}{|Q|}\,.
\end{eqnarray}
Then the range of $Z$ where both $Z$ and $Y$ are positive ($Z\geq0, Y\geq 0$) in order to have positive energy is $0<Z<3^{3/4}$.  It is easy to check that for this range of $Z$ we have 
\begin{eqnarray}
Y=\frac{m_0}{|Q|}<1 ~~~ \rightarrow ~~~ m_0< |Q| \,. 
\end{eqnarray}
Therefore, the positivity of $m_0$ and $M$ gives that  
\begin{eqnarray}\label{overcharged14}
m_0<|Q| ~~~ \mbox{and} ~~~ M_0<3^{3/4}|Q|^{-1/2} \,,
\end{eqnarray} 
so that the shell with $w=-1/4$ can be in equilibrium only when it is overcharged. 

\subsection{Dynamics of overcharged shells}

We can also study the exact dynamics as described by Eq.(\ref{israel}). 
The latter can be written for a shell with a \textit{general} equation of state as
\begin{eqnarray}
\Big(1-\frac{2m}{R}+\frac{Q^2}{R^2}+\dot R^2\Big)^{1/2}-\Big(1+\dot R^2\Big)^{1/2}=-\frac{M_0}{R^{1+2w}}\,. \label{israel2}
\end{eqnarray}
To study overcharged shells, we should consider the case of a negative $w<0$ since as we have seen above, when $w\geq 0$ ($p\geq 0$) the shell is always undercharged. \footnote{Note that if we assume that the shell can go (or start) from infinity $R\to \infty$, we get for negative $w$, $-\frac{1}{2}\leq w<0 $. Note that for this range of $w$, the potential in Eq.\eqref{potential} is bounded at $R\to \infty$ as expected.}
Taking a derivative of Eq.(\ref{israel2}) we find the equation of motion of the shell
\begin{eqnarray}
\ddot{R}=-\frac{\partial V}{\partial R}\,, \label{eom}
\end{eqnarray}
where the potential $V(R)$ is given by
\begin{eqnarray}\label{potential}
V(R)=-\frac{1}{8M_0^2}\left(\frac{M_0^2}{R^{1+2w}}-Q^2R^{2w-1}+2m R^{2w}\right)^2.
\end{eqnarray}
The potential $V$ is of the form 
\begin{eqnarray}
V(R)=-W^2 \,,
\end{eqnarray}
where the prepotential $W$ is 
\begin{eqnarray}\label{W}
W=\frac{1}{2\sqrt{2}\, M_0} \left(\frac{M_0^2}{R^{1+2w}}-Q^2R^{2w-1}+2m R^{2w}\right).
\end{eqnarray}
In other words, the dynamics is described by Eq.(\ref{eom}) subject to the initial value equation (from Eq.(\ref{israel2}))
\begin{eqnarray}
\frac{1}{2}\dot{R}^2=W^2-\frac{1}{2} \,. \label{eW}
\end{eqnarray}
The critical points of the potential are specified by solving $V'(R)=0$, which are  the zeroes of the equations
\begin{eqnarray}
W=0 \,,~~~~\partial_R W=0 \,, \label{cr}
\end{eqnarray}
which are explicitly written as 
\begin{align}
M_0^2-Q^2 R_0^{4w}+2m_0 R_0^{1+4w}&=0 \,, \label{s1}\\
M_0^2(1+2w)+Q^2(2w-1) R_0^{4w}-4m_0 w R_0^{1+4w}&=0 \,.  \label{s2}
\end{align}
The critical point specified by solving Eq.(\ref{s1}) are local global maxima  since 
\begin{eqnarray}
V''(R)\Big|_{W=0}=-|\partial_R W|^2<0 \,.
\end{eqnarray}
The second critical point at $\partial_R W=0$ is a local minimum since the potential $V\to 0$ as $R\to \infty$. 

\begin{figure}[!htbp]
\begin{subfloat}
  \centering
  \includegraphics[width=7.5cm]{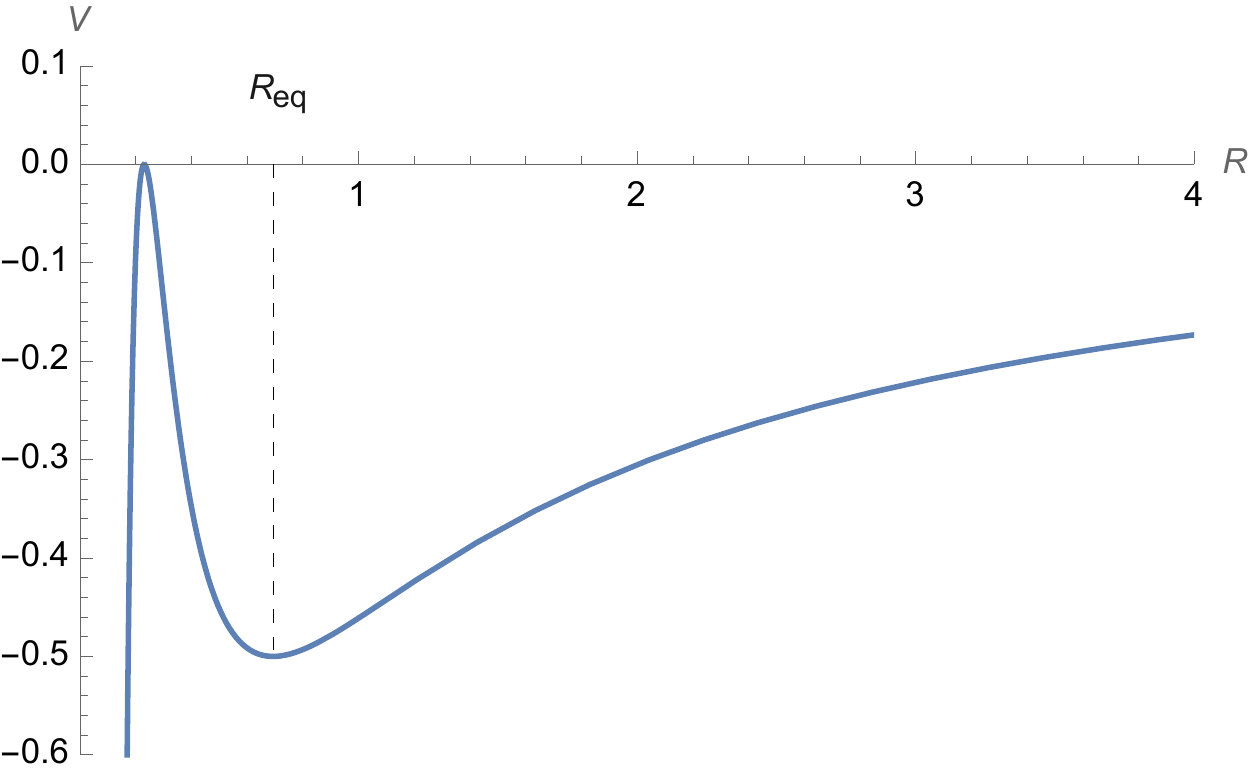}
\end{subfloat}%
\hspace{0.75cm}
\begin{subfloat}
  \centering
  \includegraphics[width=7.5cm]{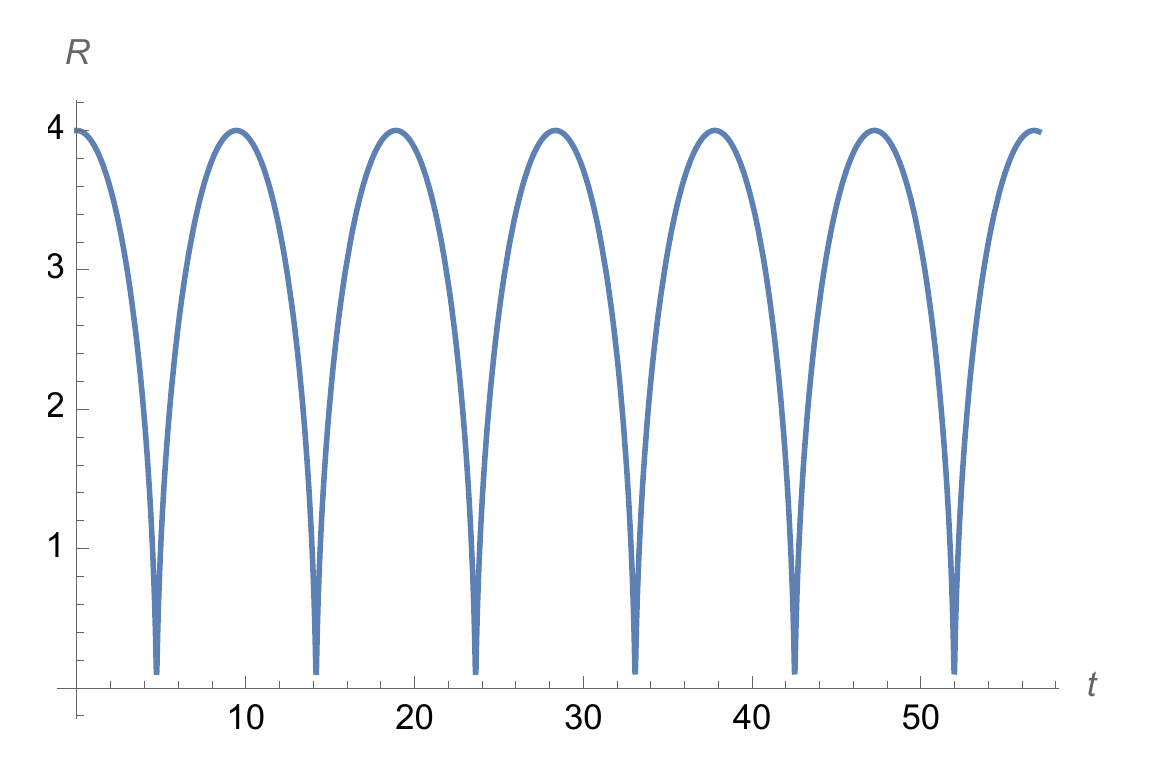}
\end{subfloat}
\caption{\label{Fig2}~ {\it Left panel}: The potential $V$ of the overcharged shell with $w=-1/4$ discussed in subsection \eqref{w14}. We choose $M_0=3^{1/2}|Q|^{-1/2}$ and $Q=1$. For these values the conditions in Eq.\eqref{overcharged14} are satisfied and the total energy is calculated to be $m_0= 3/2(3^{1/3}-1)\approx 0.66$. The potential gets the analytic form $V=-\frac{(1-3\times 3^{1/3}R)^2}{24 R^3}$ and equilibrium radius at $R_{eq}\approx 0.69$. The latter is an asymptotically increasing function for large values of $R$.  {\it Right panel}: The equations of motions of the shell around the equilibrium in terms of time obtained using Eq.\eqref{eom} and Eq.\eqref{mt}. We plot the oscillatory behavior for the initial values $R(0)=0$ and $\dot{R}(0)=4.$}
\end{figure}

The potential is plotted in Fig \ref{Fig2}. The maximum distance $R_{\rm max}$ where the  shell can be is, for initial velocity $\dot{R}_{\rm in}=0$,  the largest root of (from Eq.(\ref{eW})) 
\begin{eqnarray}
W\big(R_{\rm max}\big)^2=\frac{1}{2}\,. 
\end{eqnarray}
This value of $W$ follows also from Eq.\eqref{W} for $m\equiv m_0$, the total energy of the equilibrium. 

For illustrative purposes, we will work out explicitly the case $w=-1/4$ discussed in subsection \eqref{w14} as we can get closed expressions  in this case.
In this case, the minimum of potential is specified by  $\partial_R W=0$ from where we find that the equilibrium point $R_{\rm eq}$ is at 
\begin{eqnarray}\label{Rmin}
R_{\rm eq}=\frac{3Q^2}{2m+M_0^2}\, .
\end{eqnarray}
Comparing the equation above with $R_0$ in Eq.\eqref{R0} and solving for $m$ we find that $m\equiv m_0$ where  $m_0$ is given  in Eq.\eqref{m014}. This means that we can always replace $m$ with the total energy at the vacuum.

\subsection{Parameter space of stable charged shells}
So far we have examined in our examples the stability of undercharged shells with $w\geq 0$ and the case of overcharged shells with negative pressure for the specific value $w=-1/4$ which can be solved analytically. In this subsection we aim to find the stability regions of both over- and under-charged shells scanning the parameter space of $w$. To do so we study extensively the equilibrium/stability conditions given in Eqs.\eqref{eq2}-\eqref{eq3} and plot the regions where they are satisfied. 

As a first step we reduce the variables in these conditions using Eq.\eqref{mgeneral} and expressing the mass of the shell in the following general form
\begin{align}\label{mass2}
    M_0=R_0^{2w}\left(R_0\pm\sqrt{Q^2-2m_0R_0+R_0^2}\right)\,.
\end{align}
It is useful to introduce the following normalized variables of charge and radius in units of $m_0$
\begin{align}
    \tilde{Q}=\frac{\vert Q\vert}{m_0}\,,\,\,\,\,\,\,\tilde{R}=\frac{R_0}{m_0}\,,
\end{align}
and then we re-express the Eq.\eqref{mass2} and the stability conditions in Eqs.\eqref{eq2}-\eqref{eq3} in terms of them
\begin{align}
& m_0(m_0\tilde{R})^{2w}\left(\tilde{R}\pm\sqrt{\tilde{Q}^2+\tilde{R}(\tilde{R}-2)}\right)=M_0\, , \hspace{-1cm}\label{redef1}\\
&(1-\tilde{R})\tilde{R}-2w(\tilde{Q}^2-2\tilde{R}+\tilde{R}^2)\mp(1+2w)\tilde{R}\sqrt{\tilde{Q}^2+\tilde{R}(\tilde{R}-2)}=0\, ,\label{redef2} \\
&\tilde{R}(6w^2+5w+1)\left(-\tilde{R}\mp\sqrt{\tilde{Q}^2+\tilde{R}(\tilde{R}-2)}\right)-\tilde{Q}^2(4w^2+3w)+\tilde{R}(8w^2+6w+1)>0\, . \label{redef3}
\end{align}
In the case of charged shells stabilized outer of the external Reissner-Nordstr\"om horizon in Eq.\eqref{fR}, one should consider the horizon condition and written in terms of the new variables takes the following form
\begin{align}\label{horizon2}
\tilde{R}>1+\sqrt{1-\tilde{Q}^2} \,,\,\,\,\,\,\,\,\,\,\,\, 0\leq \tilde{Q}\leq 1\,.
\end{align}
The solutions of the stability conditions are complicated analytic expressions and we solve them numerically. However, we demonstrate the following characteristic cases with  :
\begin{itemize}
    \item \textit{w $\geq 0$}: there are no overcharged shells in this case, shown in \eqref{subsection31}.
    \item \textit{$w=0$} : There is no stable equilibrium for any charge.
    \item \textit{$w=-1/4$} : Stable equilibrium for $\tilde{Q}>1$, presented in subsection \eqref{w14}.
    \item \textit{$w=-1/2$} : Stable equilibrium for $\tilde{Q}^2=\tilde{R}$ and $\tilde{Q}\geq 1$. Minimizing the total energy including $M_0$ one can find explicitly that 
    $$R_0=Q/\sqrt{M_0(2-M_0)}\,,$$ 
    and then $$m_0=Q\sqrt{M_0(2-M_0)}\,.$$ This quantity is bounded for $0<M_0< 2$ and gives overcharged $\tilde{Q}>1$ and the extremal case $\tilde{Q}=1$ for $M_0=1$. 
    \item \textit{Extremal case $\tilde{Q}=1$} : Stable equilibrium for $-1<w<0$. For $w=-1$ the stability conditions are not satisfied.
\end{itemize}
Considering this inequality and the previous conditions, we find the parameter space where charged shells are stable as shown on the left panel of the Fig.\ref{QvsW}:
\begin{figure}[H]
\begin{subfloat}
  \centering
  \includegraphics[width=7.5cm]{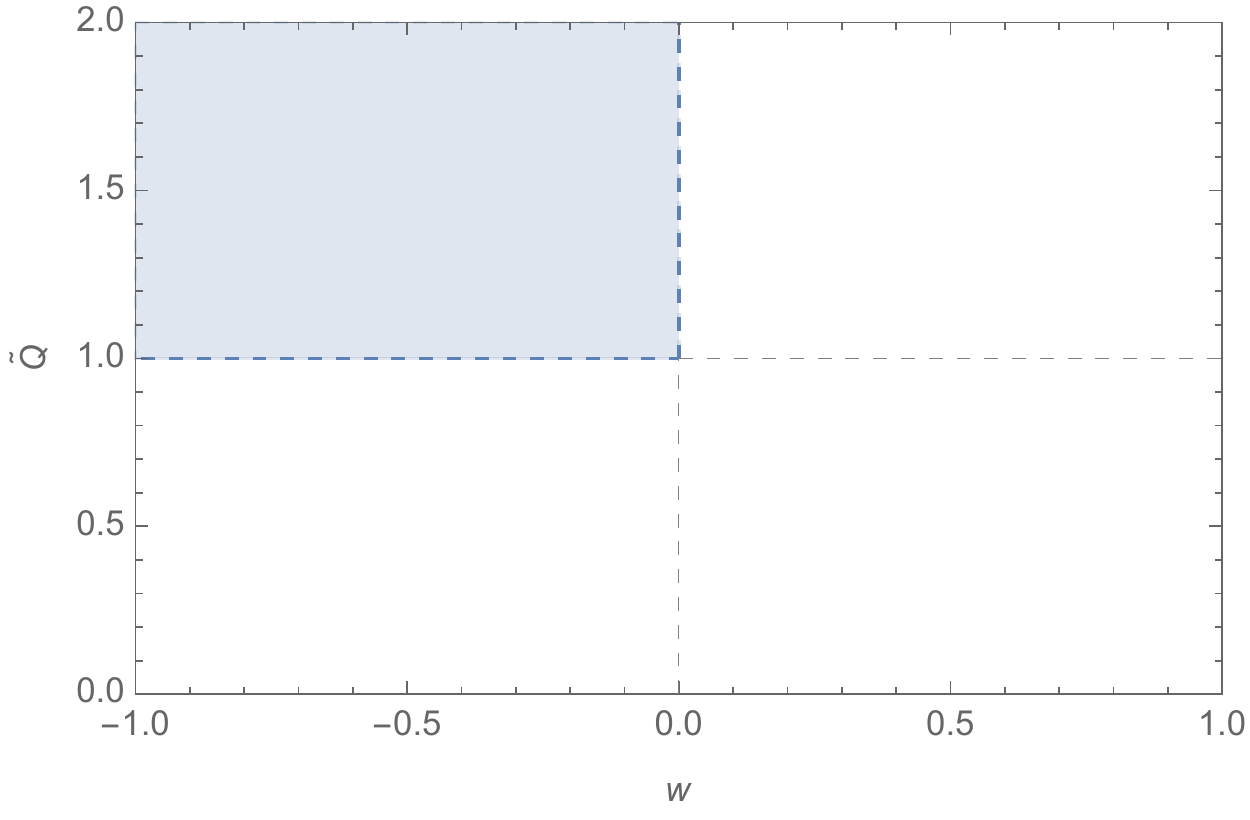}
\end{subfloat}%
\hspace{0.75cm}
\begin{subfloat}
  \centering
  \includegraphics[width=7.5cm]{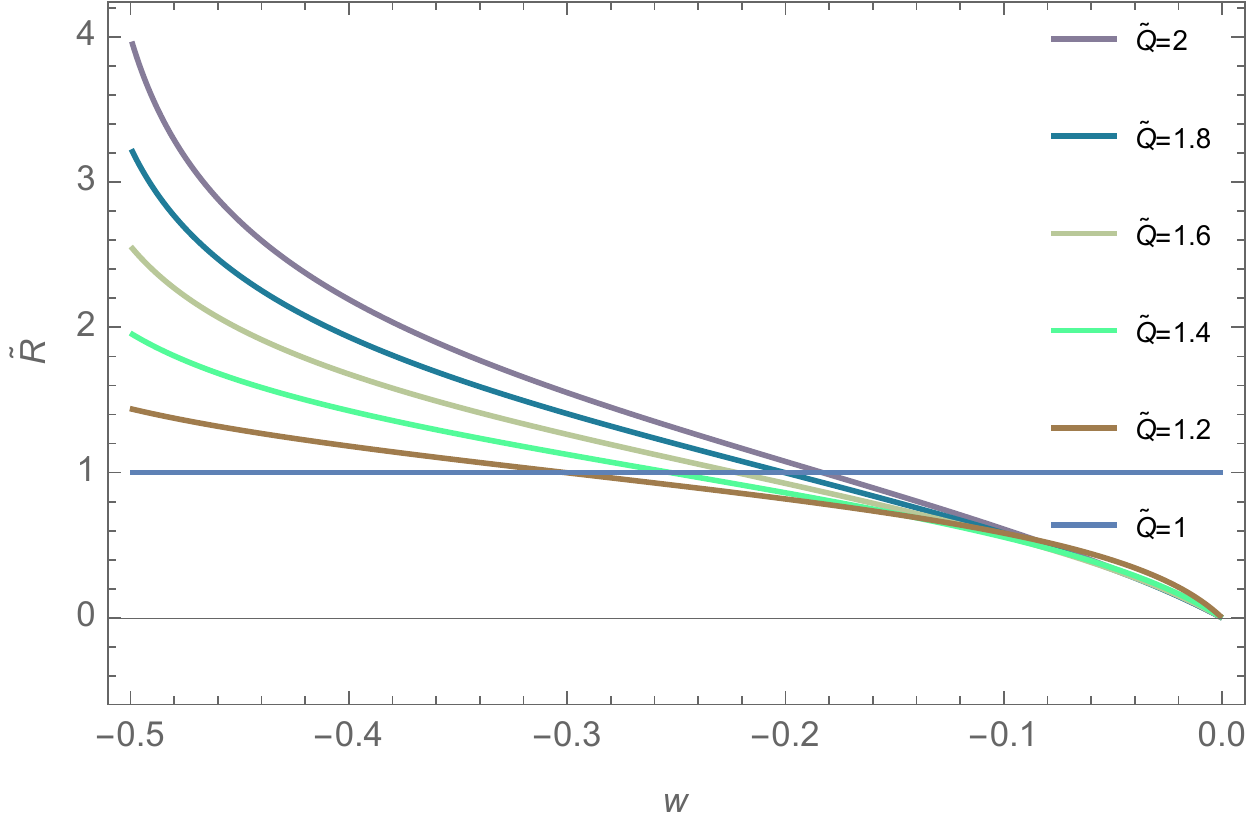}
\end{subfloat}
\caption{\label{QvsW}~ {\it Left panel} : The blue region stands for the parameter space where shells with stable equilibrium exist $\partial^2 m_0 /\partial R_0^2 >0$. Stable shells are always overcharged $\tilde{Q}>1$ with $w<0$.  {\it Right panel} : The parameter space of $\tilde{R}$ for fixed values of $\tilde{Q}$.} 
\end{figure} 

\section{Trapped spacetime modes}

Perturbations of the Reissner-Nordstr\"om spacetime has been introduced in \cite{CS} and generalized to include also a cosmological constant  in \cite{MM,Berti:2003ud}. In particular, the radial part of the  axial electromagnetic and gravitational  perturbations are described by the equation in tortoise coordinates $\text{d}R_*\equiv\text{d}R/f(R)$
\begin{align}
    \frac{d^2}{dR^2_*}Z_i^>+\left(k^2-V_i^>(R)\right)Z_i^>=0 \,, ~~~~i=1,2
\end{align}
where the potentials $V_1(R)$ and $V_2(R)$ correspond to the axial electromagnetic and gravitational perturbations, respectively. They are explicit given by 
\begin{align}\label{PotPerturbation}
    V_i^>(R)=f(R)\left(\frac{l(l+1)}{R^2}-\frac{q_j}{R^3}+\frac{4Q^2}{R^4}\right) \,,
\end{align}
where $l$ is the spherical harmonic index and 
\begin{align}
    q_{i,j}&=3m\pm \sqrt{9m^2+4(l-1)(l+2)Q^2} \,,\quad\quad i,j=1,2 \quad i\neq j\, .
\end{align}

\begin{figure}[!htbp]
\begin{subfloat}
  \centering
  \includegraphics[width=0.45\linewidth]{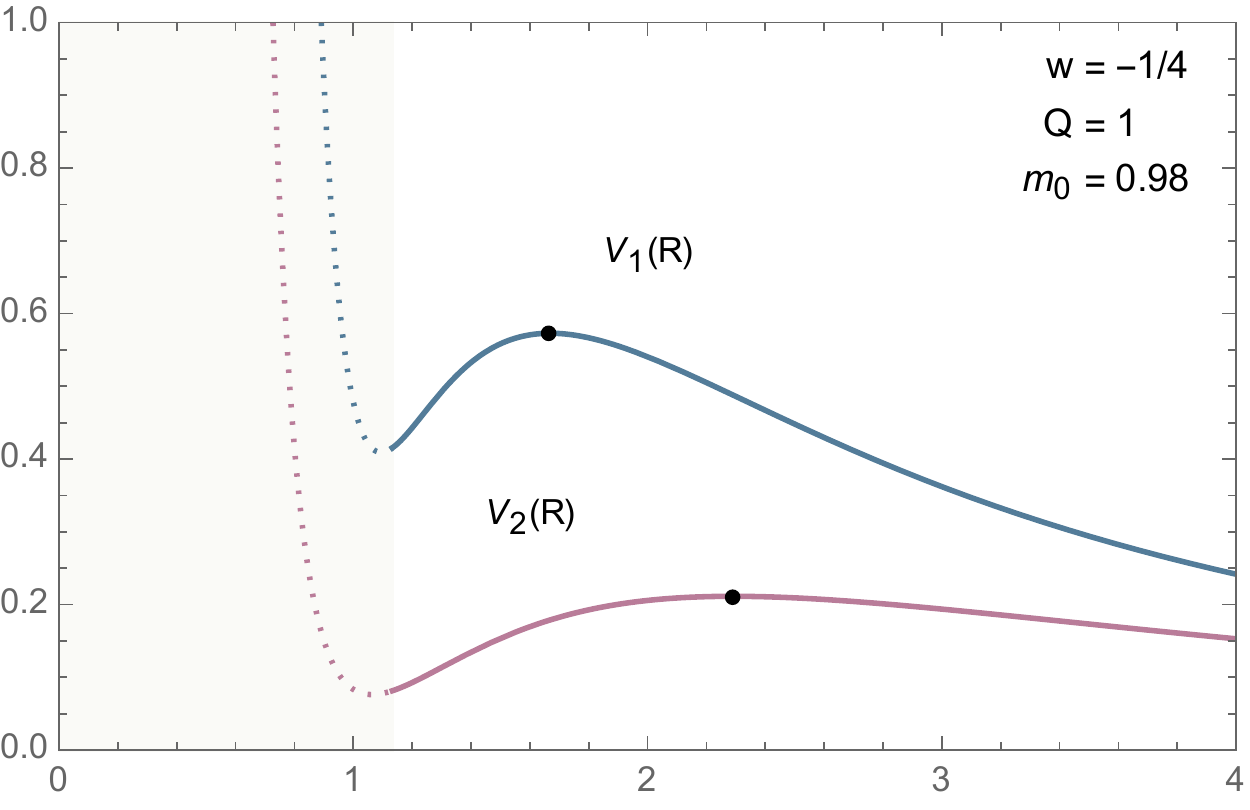}
\end{subfloat}
\begin{subfloat}
  \centering
  \includegraphics[width=0.45\linewidth]{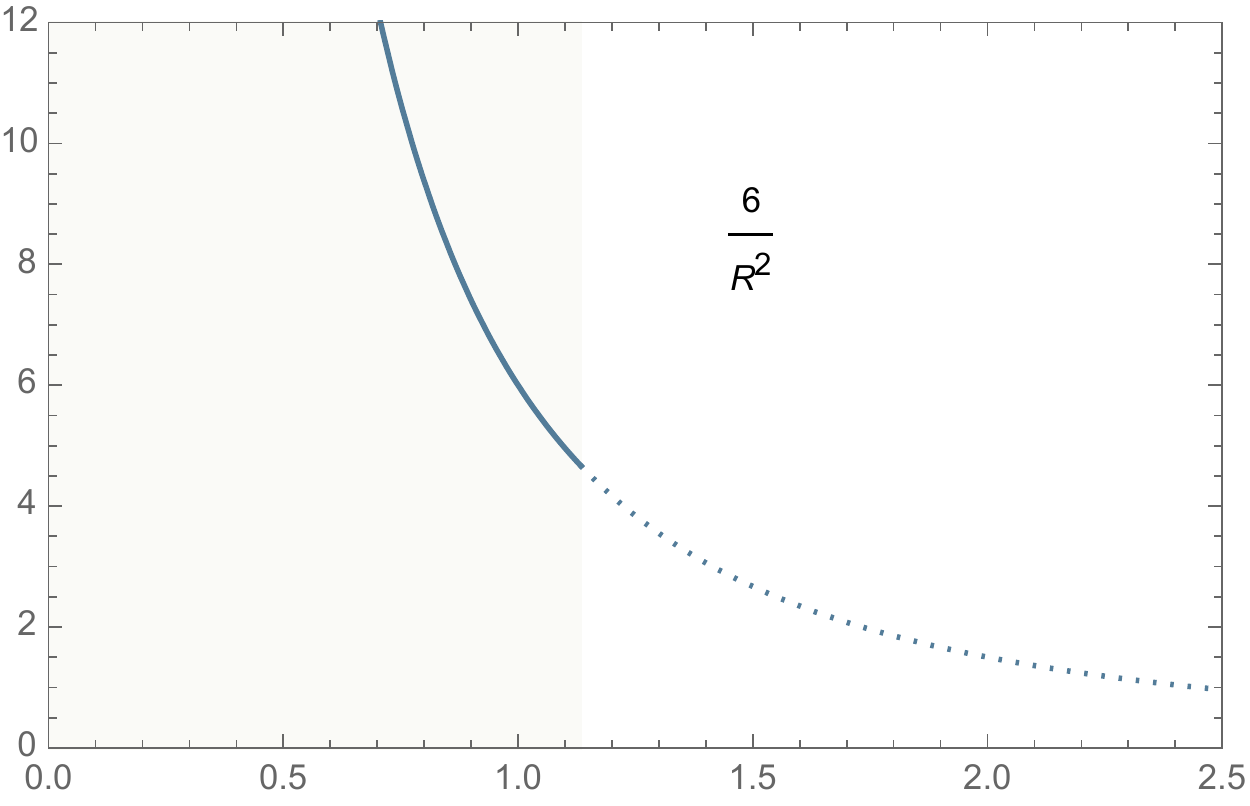}
\end{subfloat}
\begin{subfloat}
  \centering
  \includegraphics[width=0.45\linewidth]{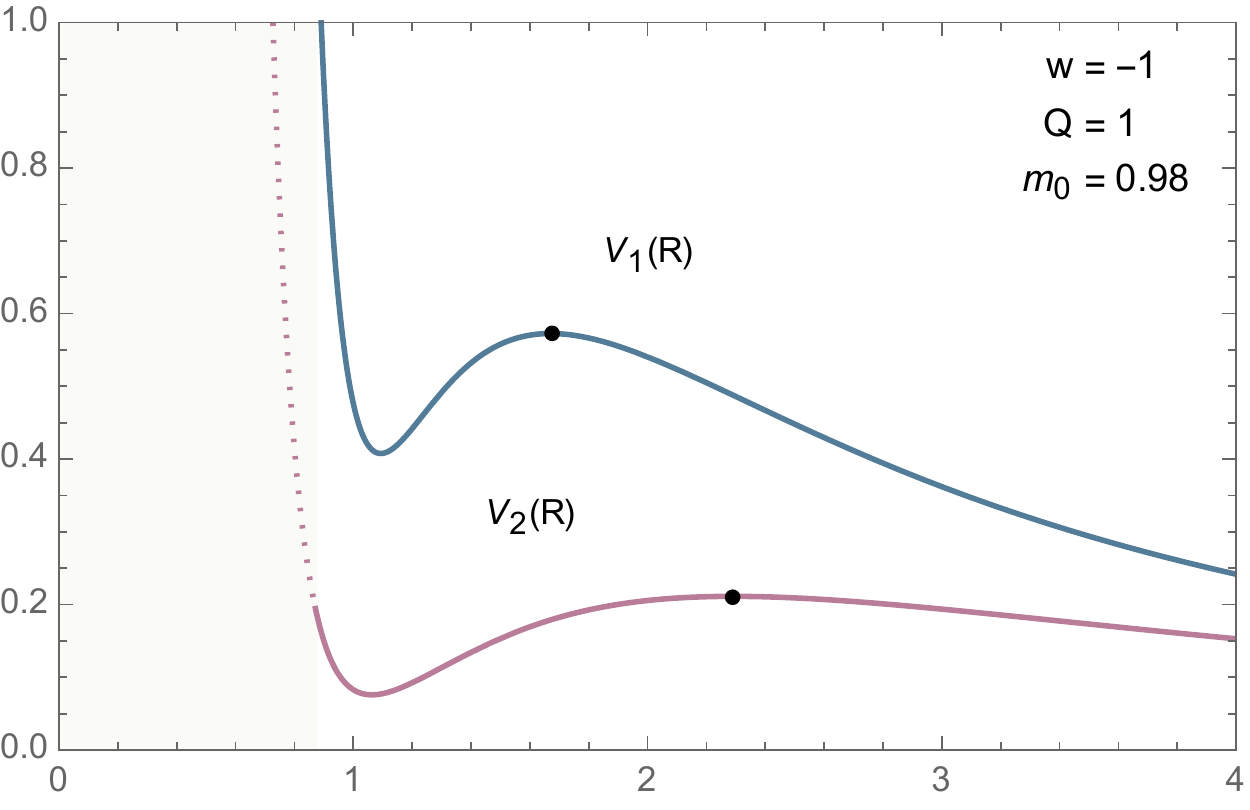}
\end{subfloat}
\hspace{1.4cm}
\begin{subfloat}
  \centering
  \includegraphics[width=0.45\linewidth]{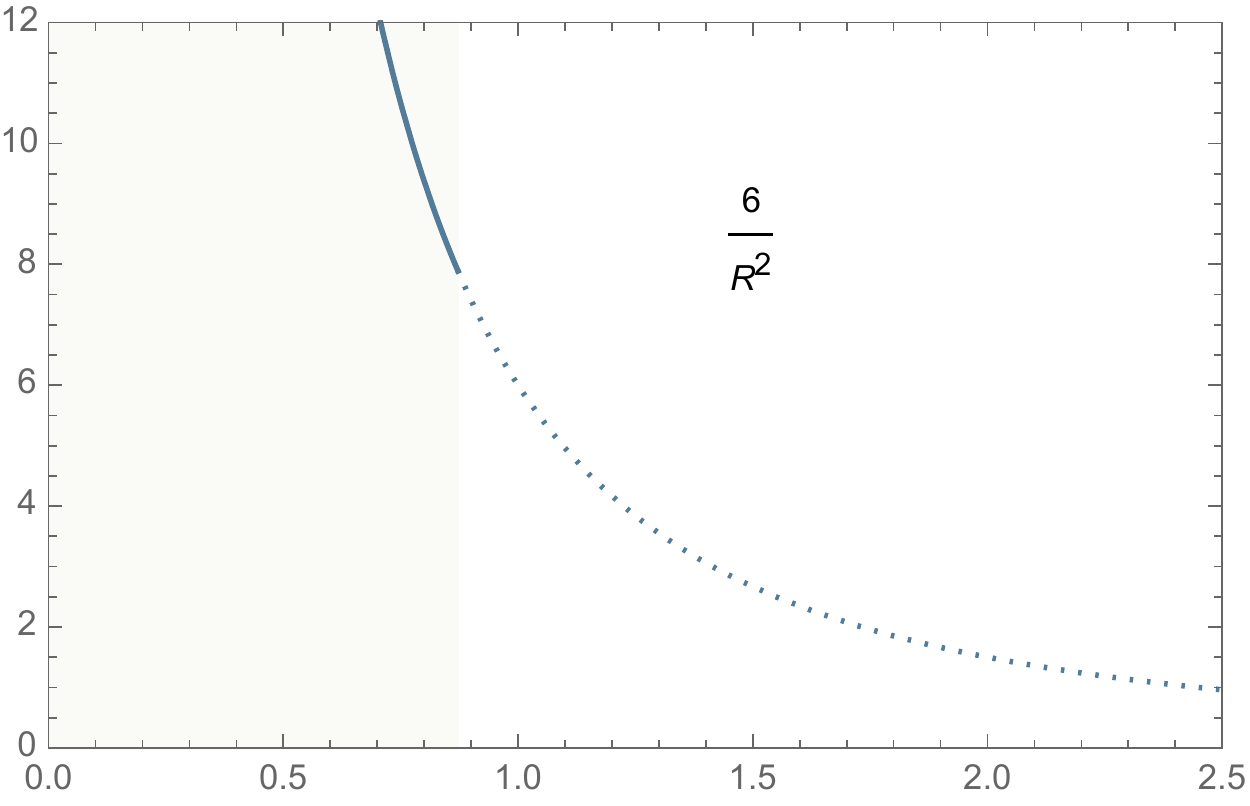}
\end{subfloat}
\caption{\label{fig3} We plot overcharged RN black holes with $m_0\approx 0.98$ and $Q=1$ and overcharged shells at equilibrium. The singularity is hidden by the overcharged shell while the shaded region corresponds to the interior of the shell. {\it Upper panel}: Overcharged shell with tension $M_0\approx 0.829$, $w=-1/4$, stabilized at $R_{\rm eq}\approx 1.133$. {\it Lower panel}: Overcharged shell with tension $M_0\approx 0.85$, $w=-1$, stabilized at $R_{\rm eq}\approx 0.87$. The total energy and charge of the shells is $m_0\approx 0.98$ and $Q=1$ respectively. Note that for negative values $w>-1$, e.g. $w=-1/4$, the equilibrium point $R_{eq}$ approaches the local maximum of the QN modes potential.}
\end{figure}

On the other hand, the spacetime is flat  inside the shell and therefore, there are  only gravitational perturbations, the radial part of which obey the equation 
\begin{align}
    \frac{d^2}{dR^2}Z_2^<+\left(k^2-V_2^<(R)\right)Z_2^<=0 \,, 
\end{align}
where 
\begin{eqnarray}
    V_2^<(R)=\frac{l(l+1)}{R^2}.
\end{eqnarray}
The potentials $V_1^>,V_2^>$ and $V_2^<$ are shown in Figs. \ref{fig3} and \ref{fig5}.

In the following, we will consider only the  gravitational perturbations with corresponding potentials $V_2^<$ and $V_2^>$ for inside and outside the shell, respectively. It is easy to see that the potential $V_2^>$ for undercharged shells with $m\geq|Q|$ has a minimum and a  maximum, as can be seen from Figs. \ref{fig3} and \ref{fig5}. The minimum is hidden behind the outer horizon, where incoming boundary conditions can be placed to describe standard QNM of Reissner-Nordstr\"om spacetime. 

\begin{figure}[!htbp]
\begin{subfloat}
  \centering
  \includegraphics[width=7.5cm]{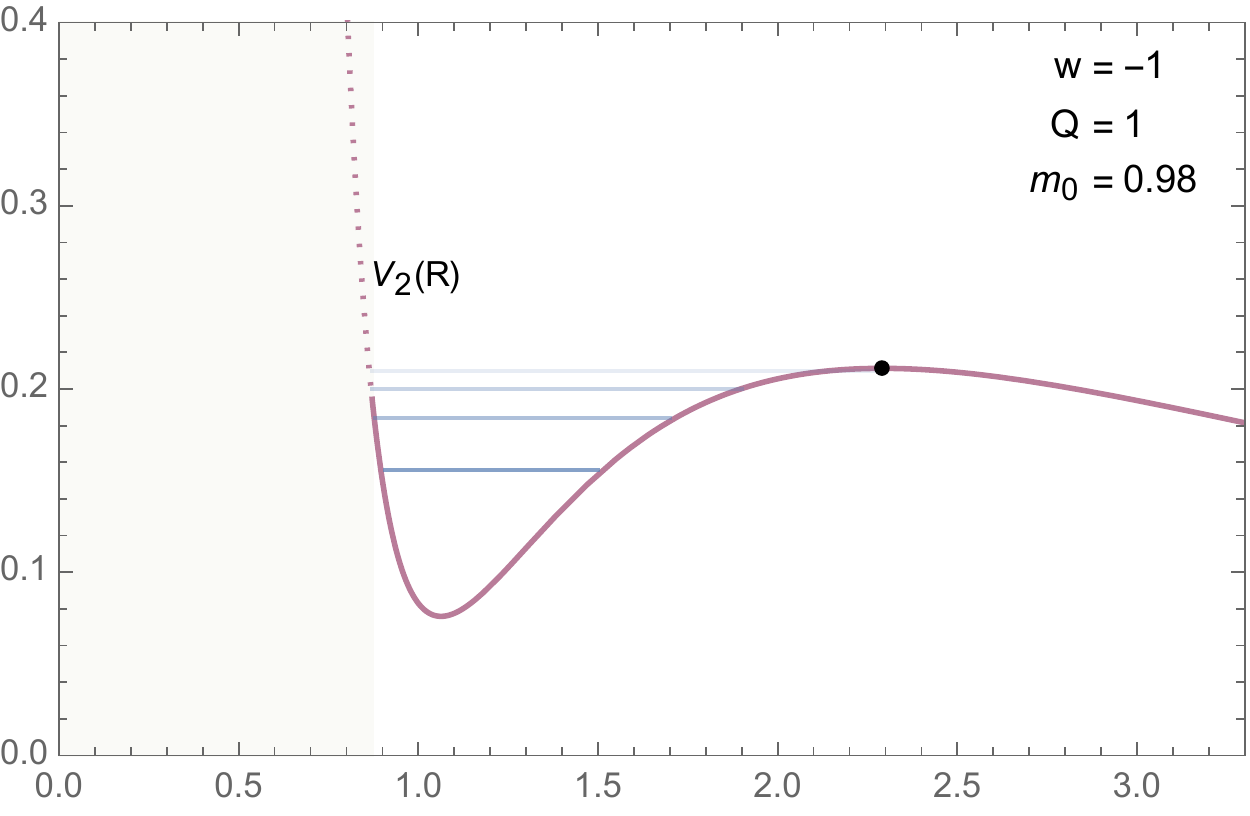}
\end{subfloat}%
\hspace{0.75cm}
\begin{subfloat}
  \centering
  \includegraphics[width=7.5cm]{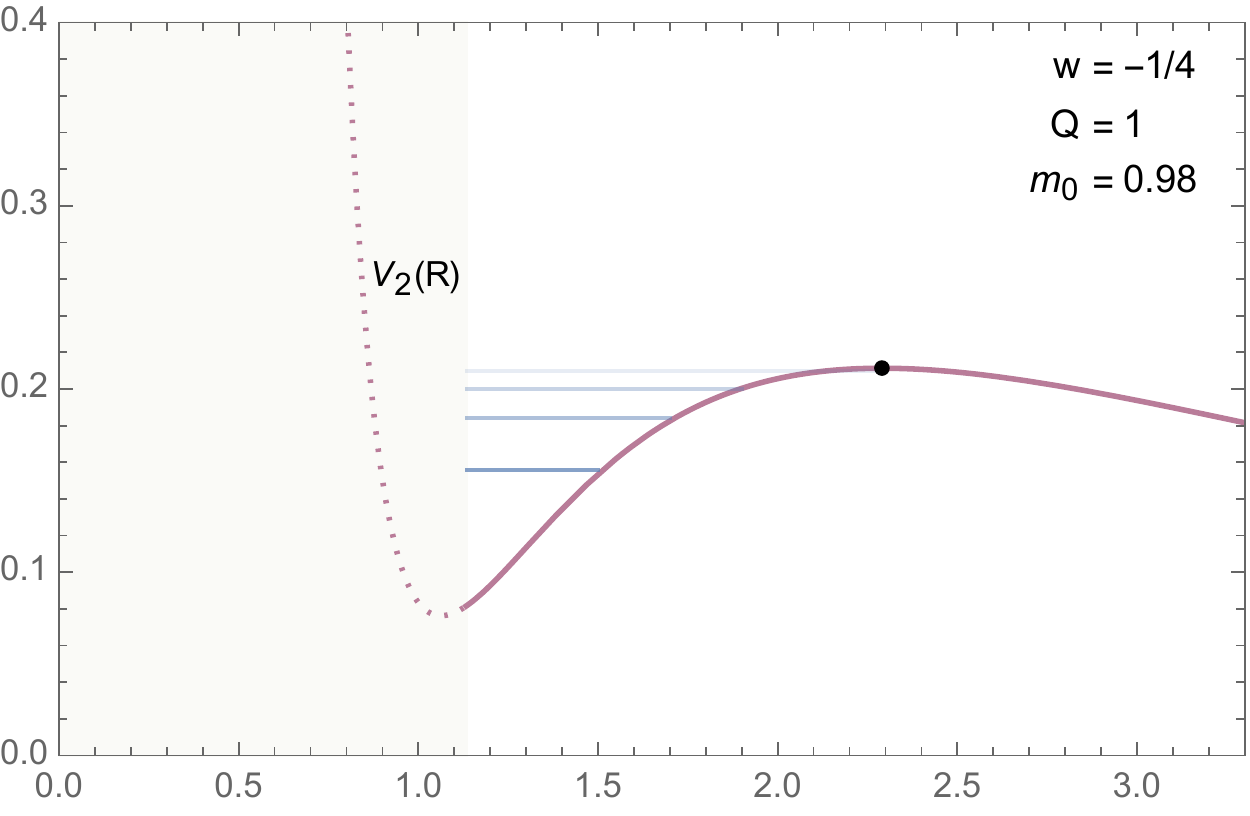}
\end{subfloat}
\caption{\label{fig4}~ {\it Left panel}: Schematic illustration of trapped spacetime modes on the exterior region of an overcharged shell with $w=-1/4$, corresponding to the upper left panel of Fig.\ref{fig3}.~ {\it Right panel}: Schematic illustration of trapped modes on the exterior region of an overcharged shell with $w=-1$, corresponding to the upper left panel of Fig.\ref{fig3}.} 
\end{figure}

\begin{figure}[!htbp]
\begin{subfloat}
  \centering
  \includegraphics[width=0.45\linewidth]{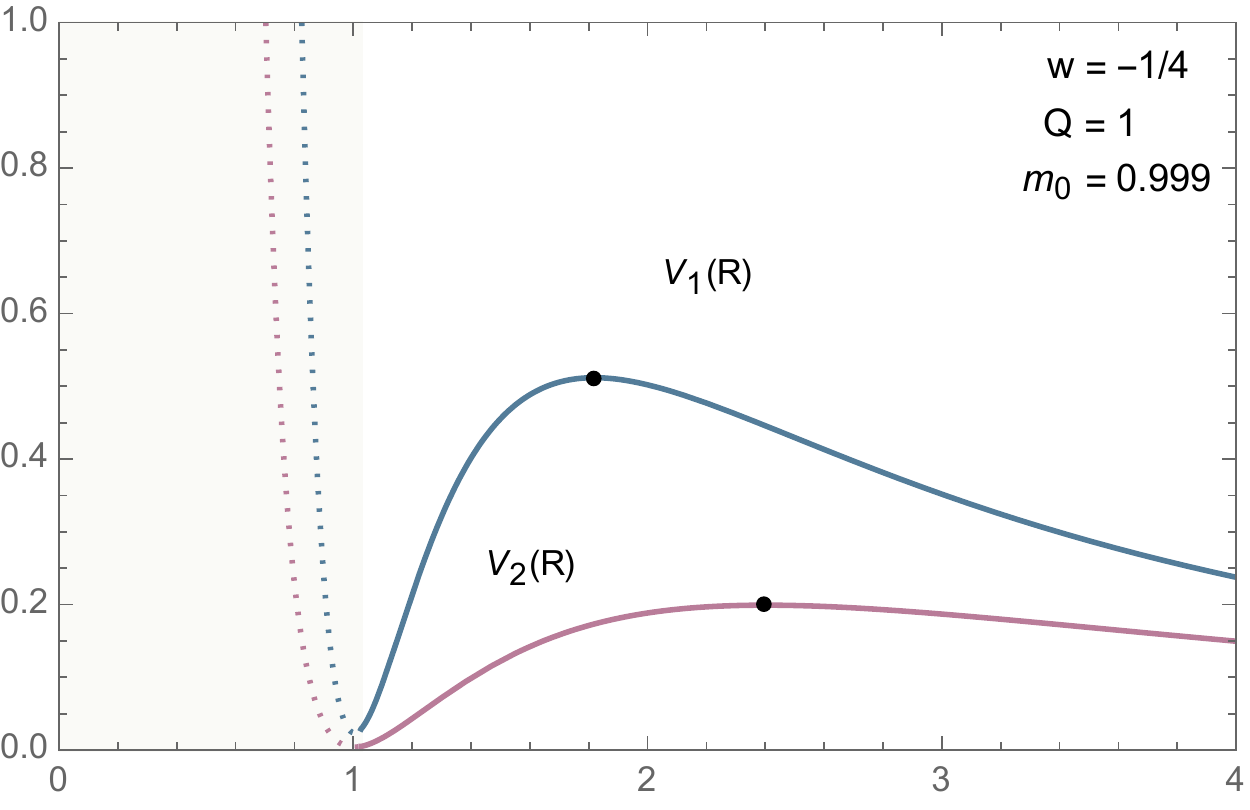}
\end{subfloat}
\begin{subfloat}
  \centering
  \includegraphics[width=0.45\linewidth]{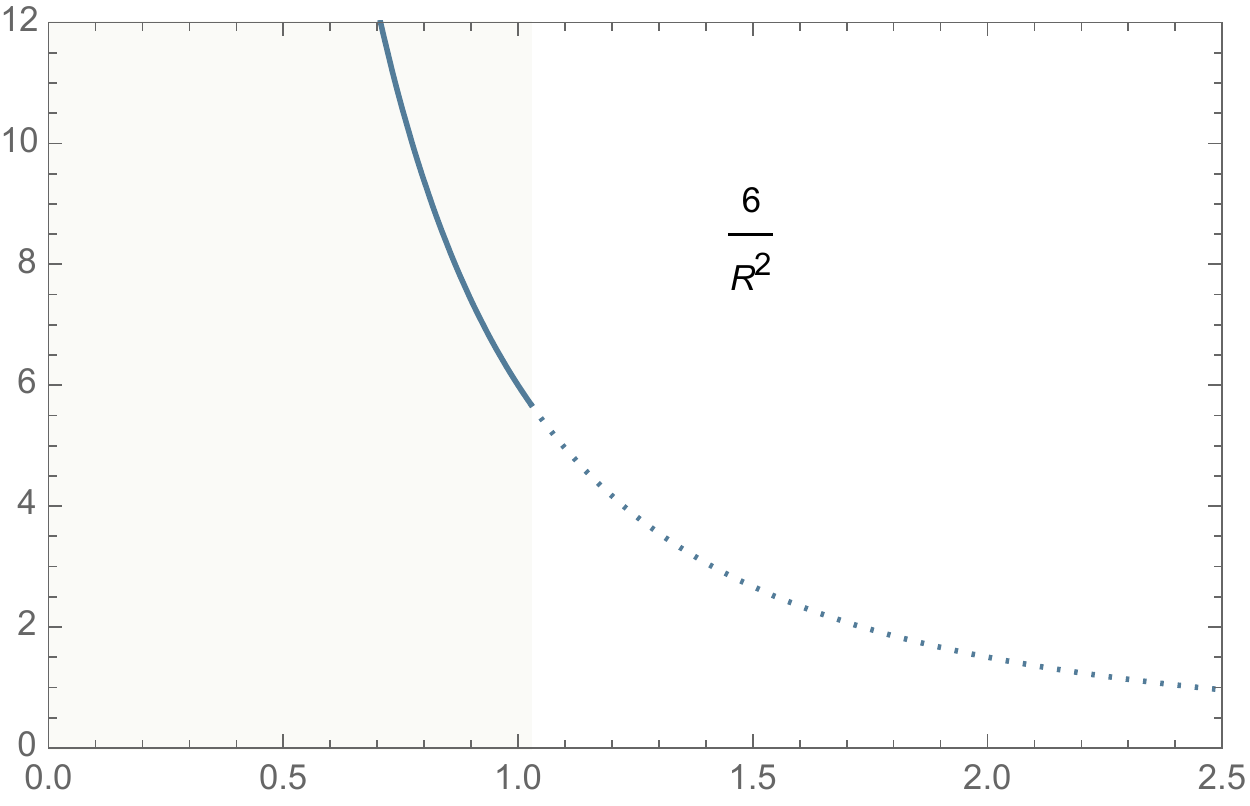}
\end{subfloat}
\begin{subfloat}
  \centering
  \includegraphics[width=0.45\linewidth]{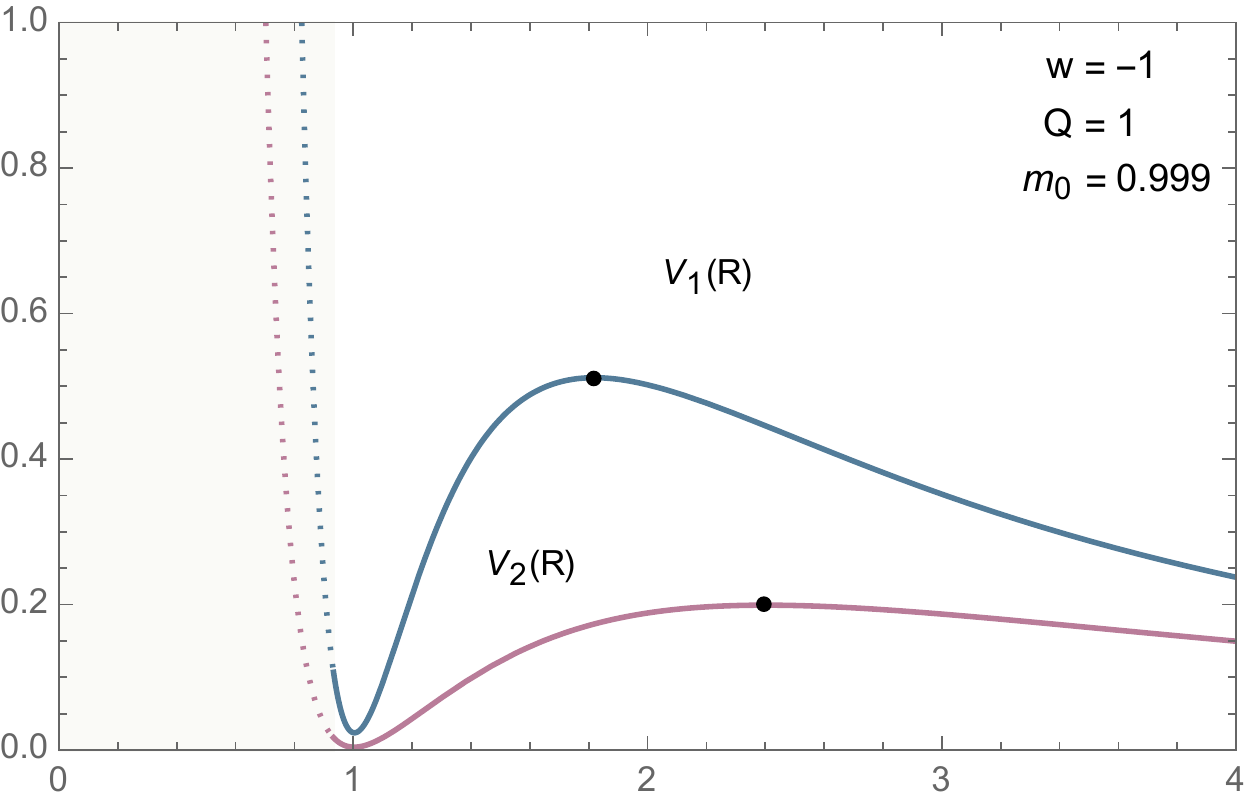}
\end{subfloat}
\hspace{1.4cm}
\begin{subfloat}
  \centering
  \includegraphics[width=0.45\linewidth]{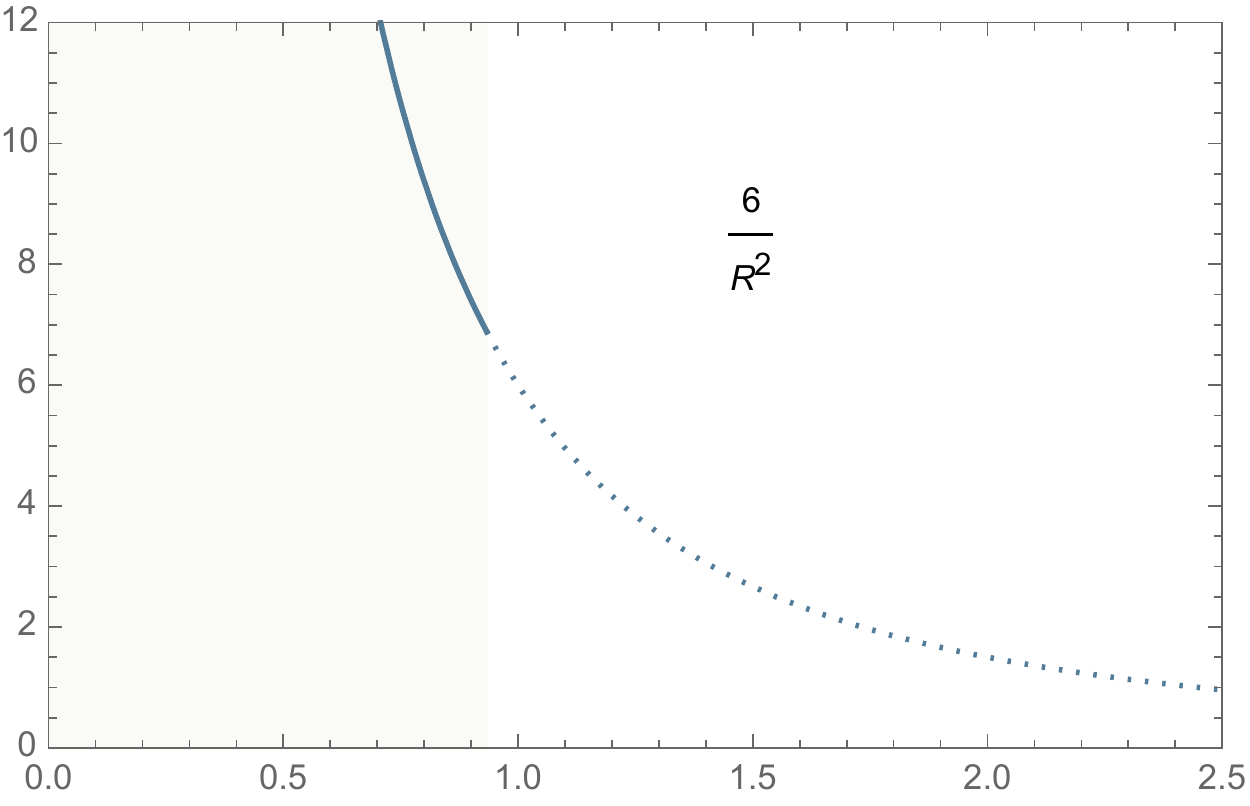}
\end{subfloat}
\caption{\label{fig5} We plot overcharged RN black holes with $m_0\approx 0.999$ and $Q=1$ and overcharged shells at equilibrium. The singularity is hidden by the overcharged shell while the shaded region corresponds to the interior of the shell. {\it Upper panel}: Overcharged shell with tension $M_0\approx 0.9614$, $w=-1/4$, stabilized at $R_{\rm eq}\approx 1.026$. {\it Lower panel}: Overcharged shell with tension $M_0\approx 0.9803$, $w=-1$, stabilized at $R_{\rm eq}\approx 0.933$. The total energy and charge of the shells is $m_0\approx 0.999$ and $Q=1$ respectively. Note that negative values $w>-1$, e.g. $w=-1/4$, the equilibrium point $R_{eq}$ approaches the local maximum of the QN modes potential.}
\end{figure}

\begin{figure}[!htbp]
\begin{subfloat}
  \centering
  \includegraphics[width=7.5cm]{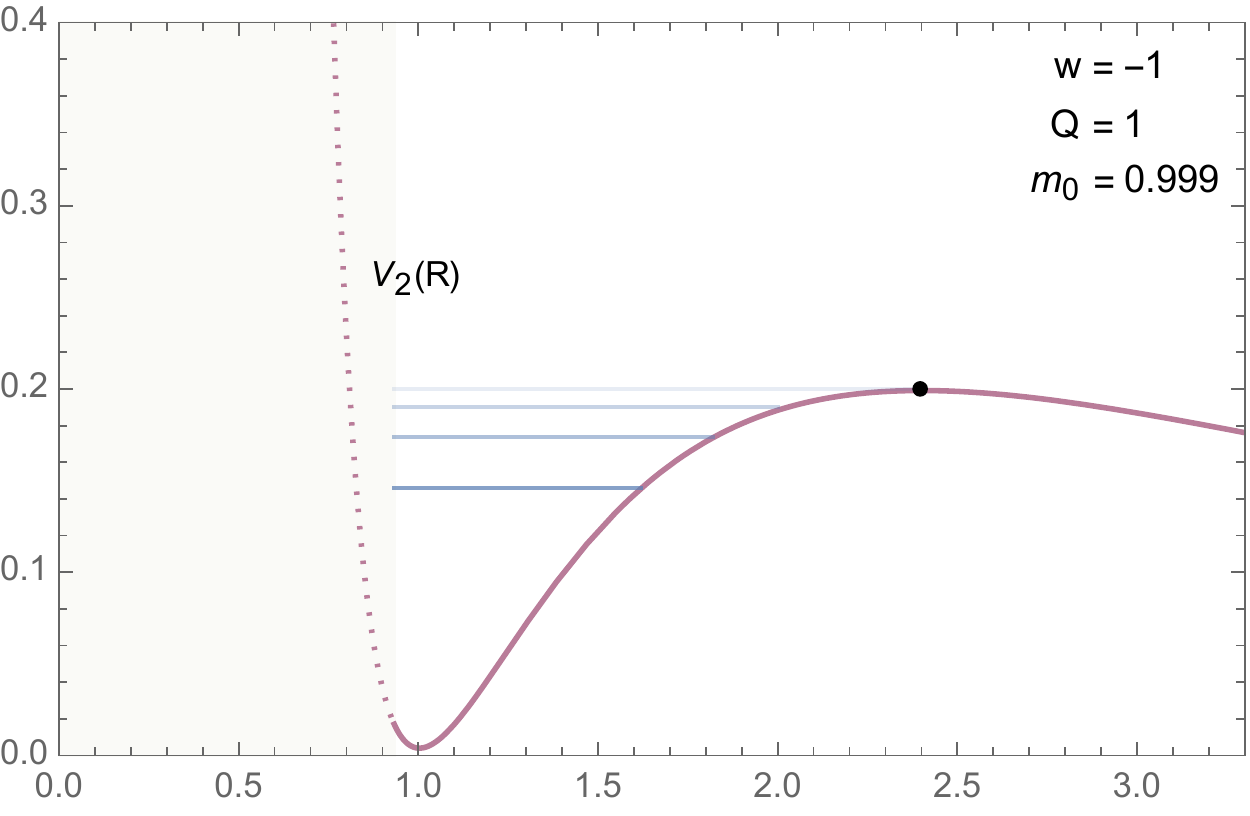}
\end{subfloat}%
\hspace{0.75cm}
\begin{subfloat}
  \centering
  \includegraphics[width=7.5cm]{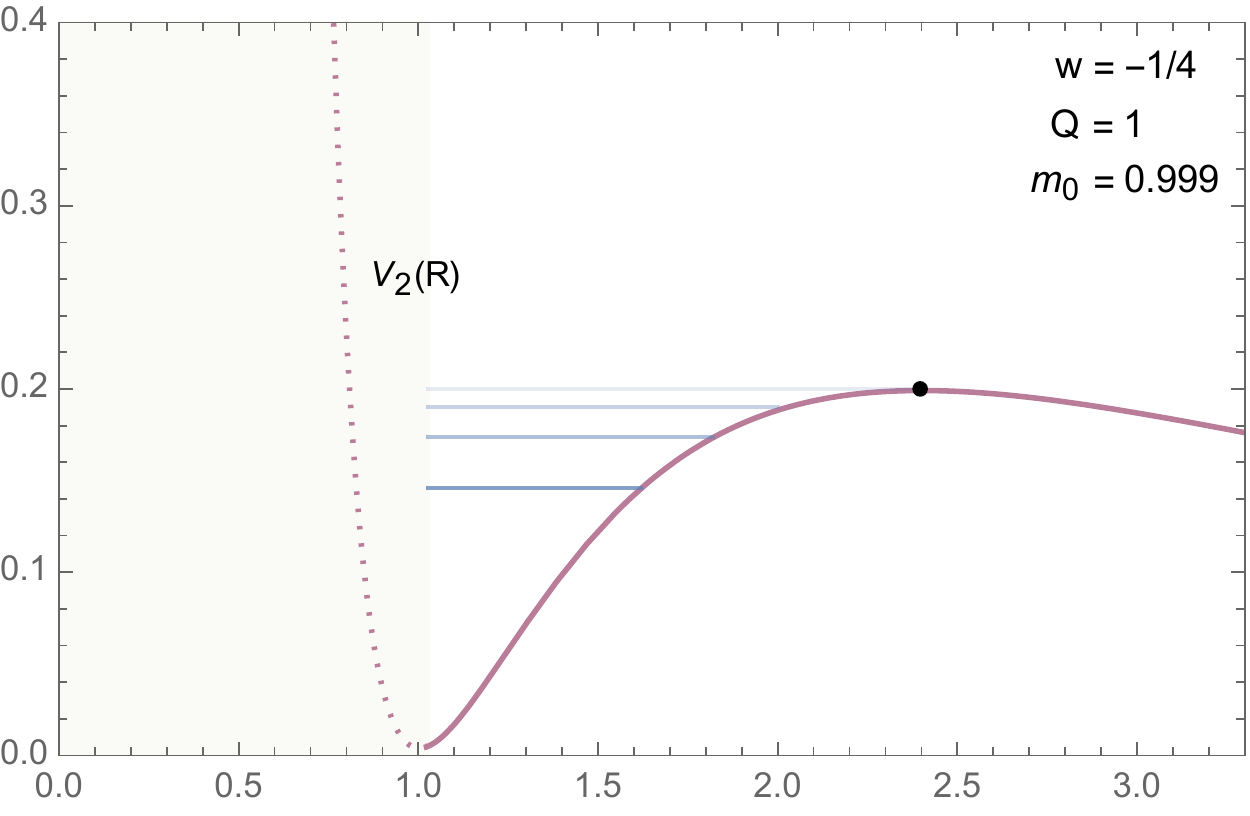}
\end{subfloat}
\caption{\label{fig6}~ {\it Left panel}: Schematic illustration of trapped spacetime modes on the exterior region of an overcharged shell with $w=-1/4$, corresponding to the Fig. \ref{fig5}.~ {\it Right panel}: Schematic illustration of trapped modes on the exterior region of an overcharged shell with $w=-1$, corresponding to the Fig. \ref{fig5}.} 
\end{figure}

\begin{figure}[!htbp]
\begin{subfloat}
  \centering
  \includegraphics[width=8.cm]{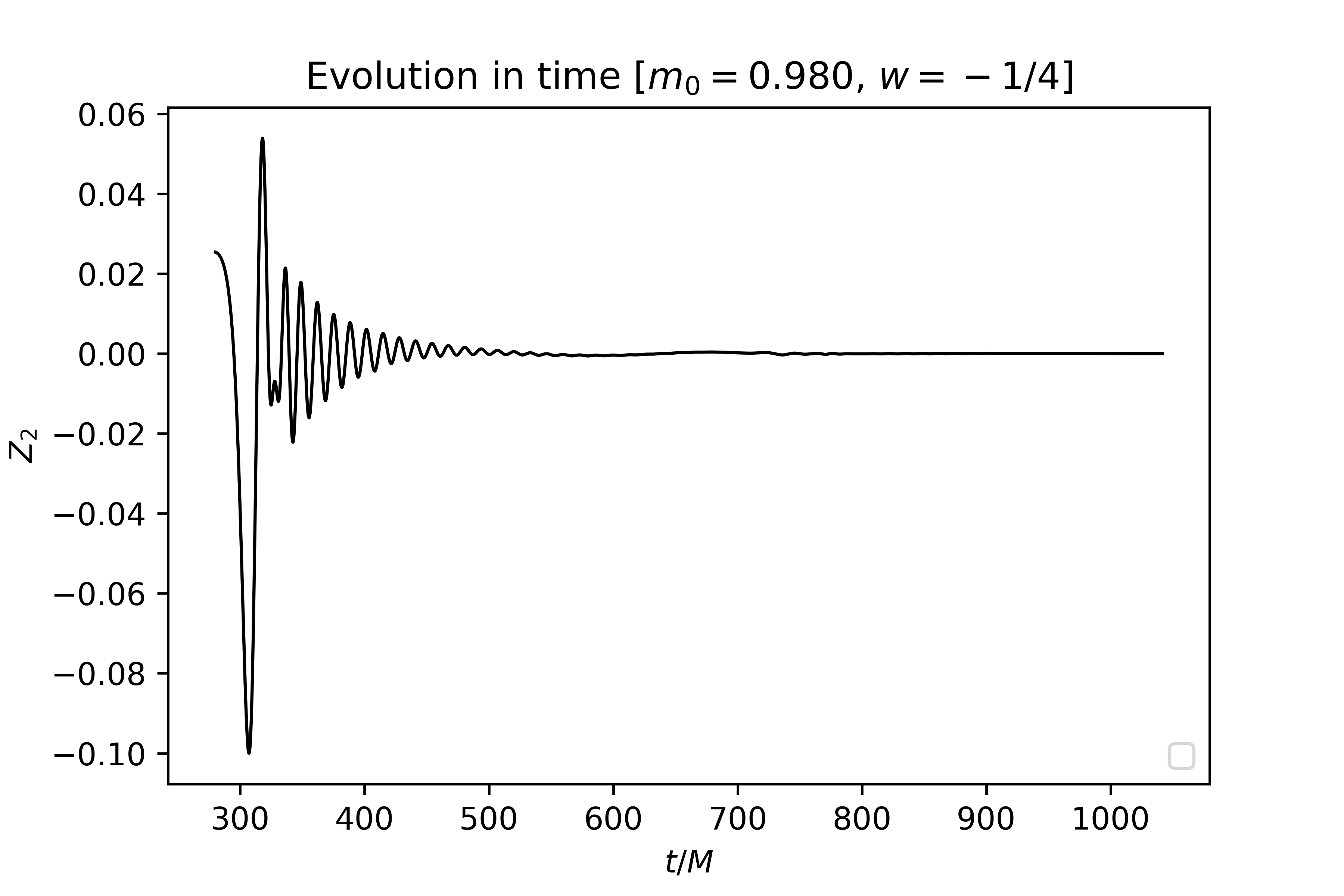}
\end{subfloat}
\begin{subfloat}
  \centering
  \includegraphics[width=8.cm]{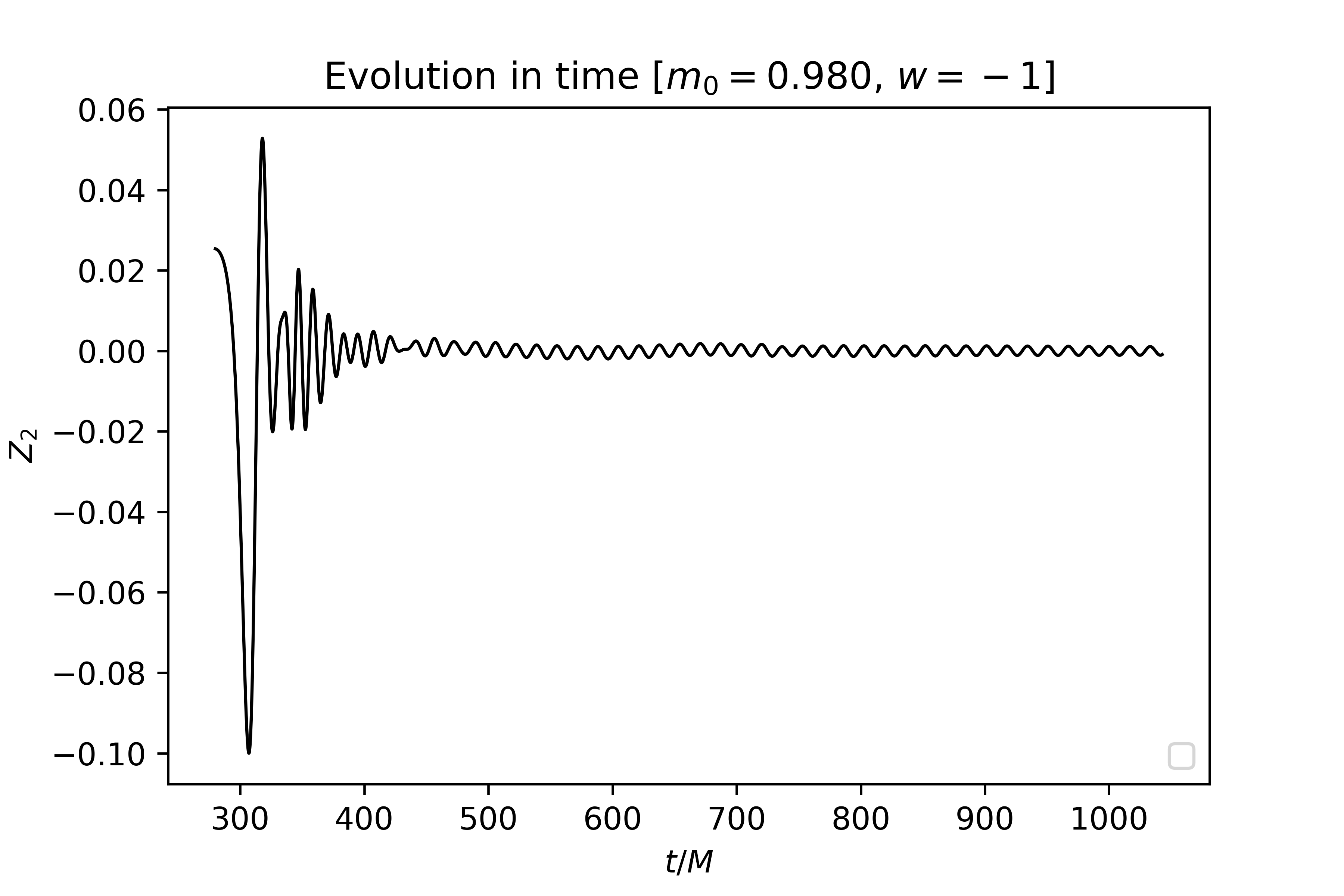}
\end{subfloat}
\begin{subfloat}
  \centering
  \includegraphics[width=8.cm]{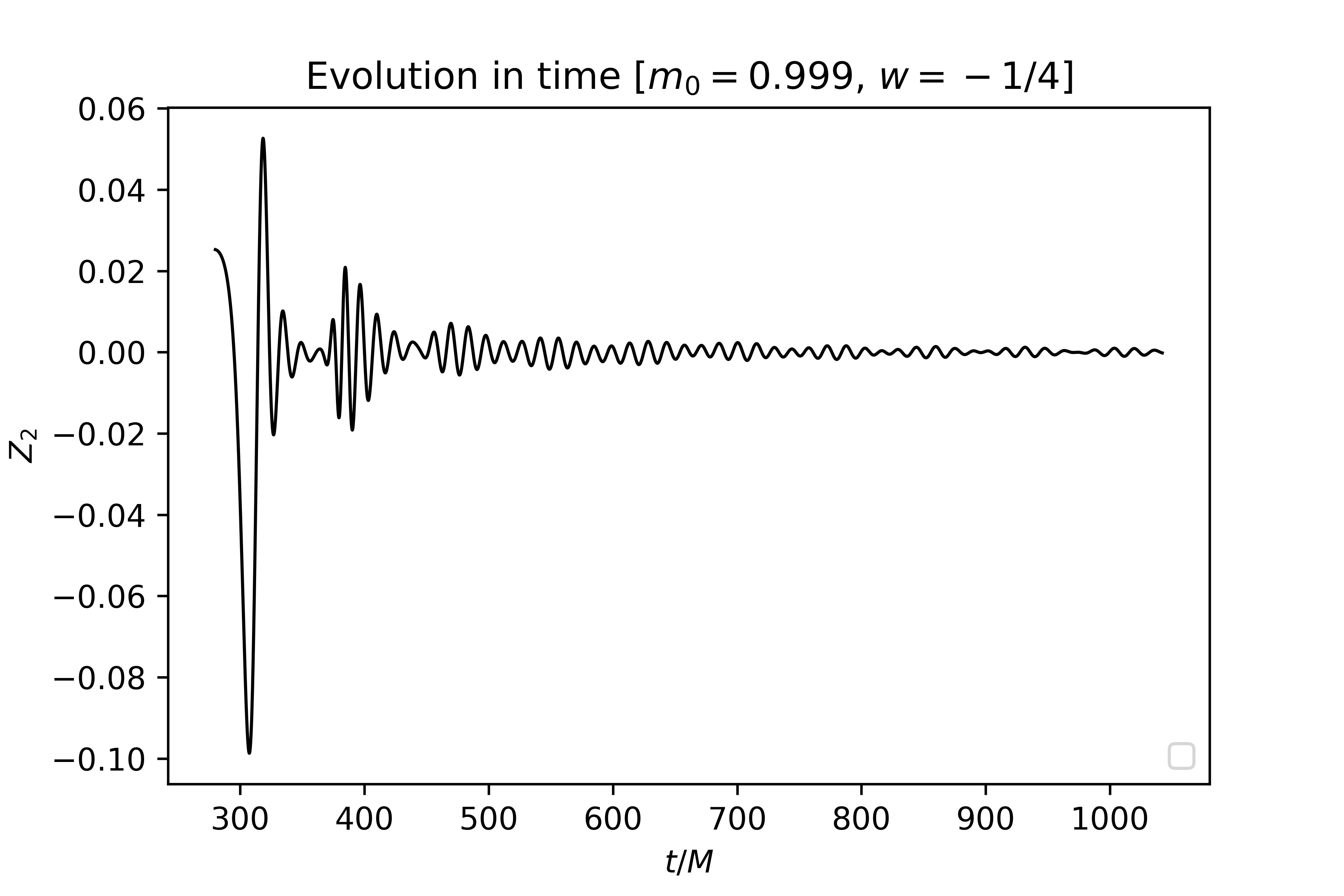}
\end{subfloat}
\begin{subfloat}
  \centering
  \includegraphics[width=8.cm]{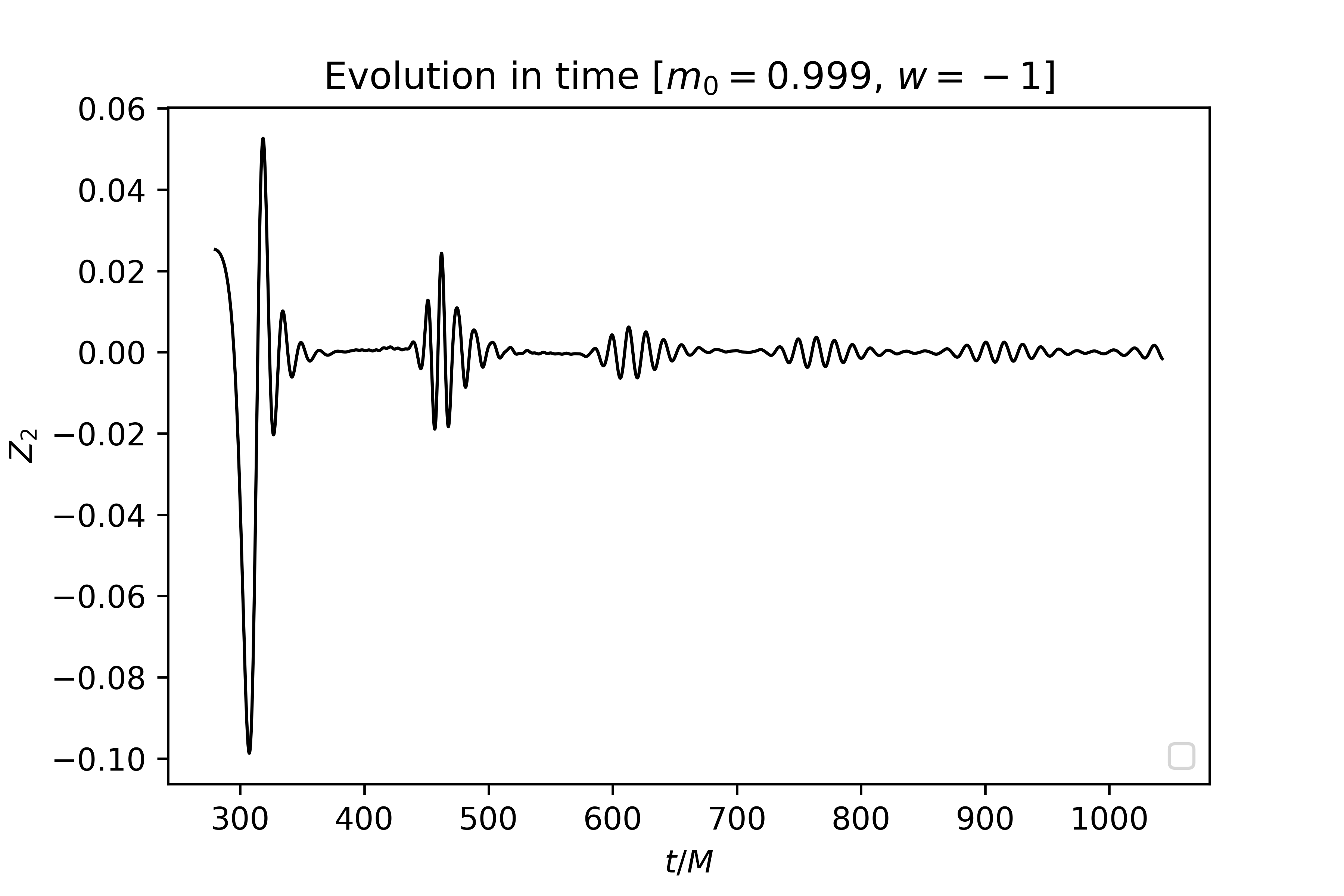}
\end{subfloat}
\caption{\label{fig7} {\it Upper panels}: The emerging signal for $m_0=0.980$ and $R_{\rm eq}\approx 1.133$ ($w=-1/4$) on the left and  $m_0=0.980$ and $R_{\rm eq}\approx 0.87$ ($w=-1$) on the right. {\it Lower panels}: The emerging signal for $m_0=0.999$ nd $R_{\rm eq}\approx 1.026$ ($w=-1/4$) on the left and  $m_0=0.999$ and $R_{\rm eq}\approx 0.933$ ($w=-1$) on the right. In all four panels the initial QNMs due to scattering on the potential barrier is followed by ``trapped'' modes. The number of the trapped modes and the energy that they carry depend on the triplet ($m_0, R_{\rm eq},w$) which define the depth and the width of the potential well. ~ }
\end{figure}

For overcharged shells with $m<|Q|$ the situation is different. In particular, for $m/|Q|\lesssim 0.912$ the potential $V_2^>$ has no extrema. In the opposite case, for $0.912\lesssim m/|Q|<1$, the potential $V_2^>$ develops a minimum at $R_{\rm min}$ and a maximum at $R_{\rm max}$ with $R_{\rm min}<R_{\rm max}$ as in Figs. \ref{fig3} and \ref{fig4}. It turns out that the shell is stabilized at a distance $R_{\rm eq}<R_{\rm max}$. Therefore, there are trapped spacetime modes \cite{kk1} between the position of the shell $R_{\rm eq}$ and the maximum of the potential at $R_{\rm max}$ as illustrated in Fig. \ref{fig6}. They can be long-lived depending on  the depth of the potential well. Trapped modes have first been studied 
for the axial stellar perturbations in \cite{cc1,KK1994}. It turns out that trapped modes exist whenever there are horizonless objects confined in the interior of the potential barrier. This is the case for example of  ultra-compact stars 
and gravastars \cite{MM2001,MM2023}
studied in \cite{TSM1999,FK2000,CFP2016,ADA2017,kk2,kk3,WA2018,MTBP2019}, where the spectrum of the perturbations, including the trapped modes, has been determined. 
If such horizonless ultra-compact objects  formed in nature once they involved in dynamical processes (binary mergers, collapse, excitations by other bodies) will produced a characteristic signals (echoes) as in Fig. \ref{fig7}. A detailed analysis of the signals will provide details and reveal the nature of the emitting object \cite{kk2,kk3,WA2018,MTBP2019}.  

\pagebreak




\begin{figure}[!htbp]
\begin{subfloat}
  \centering
  \includegraphics[width=7.5cm]{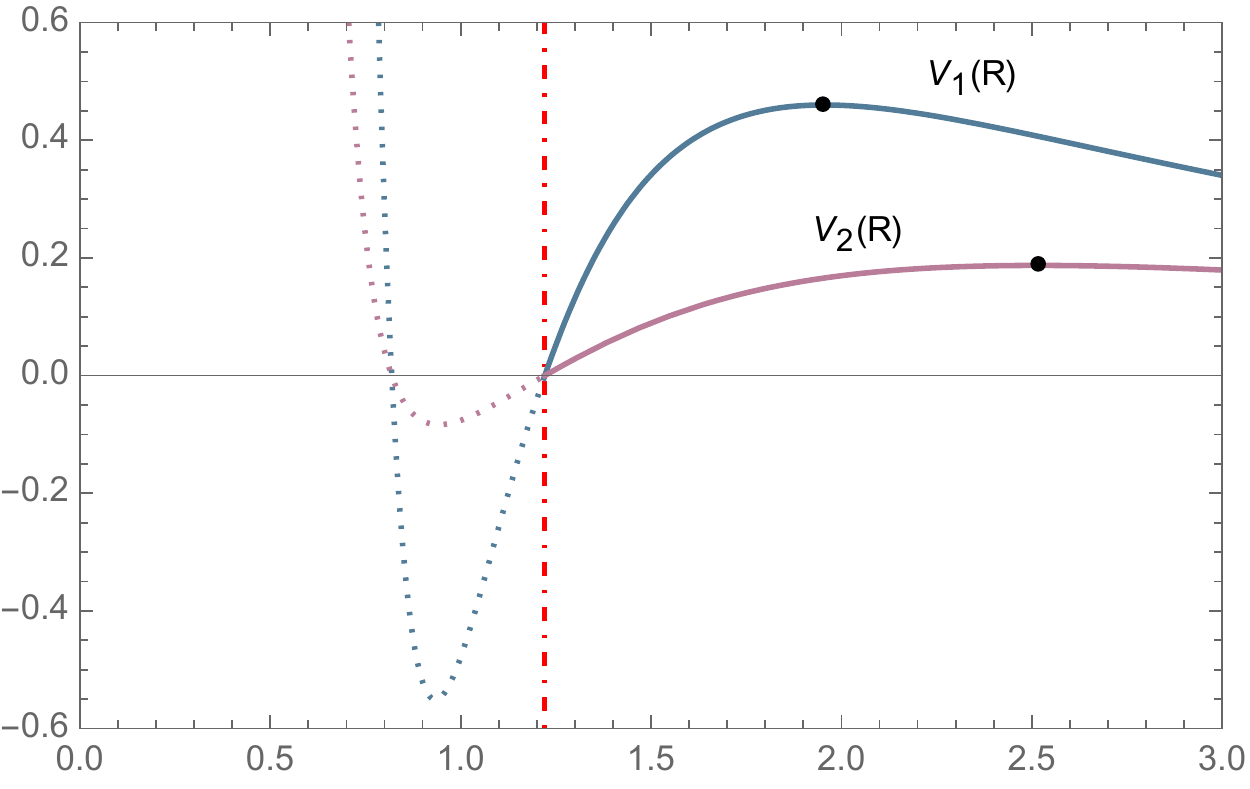}
\end{subfloat}%
\hspace{0.75cm}
\begin{subfloat}
  \centering
  \includegraphics[width=7.5cm]{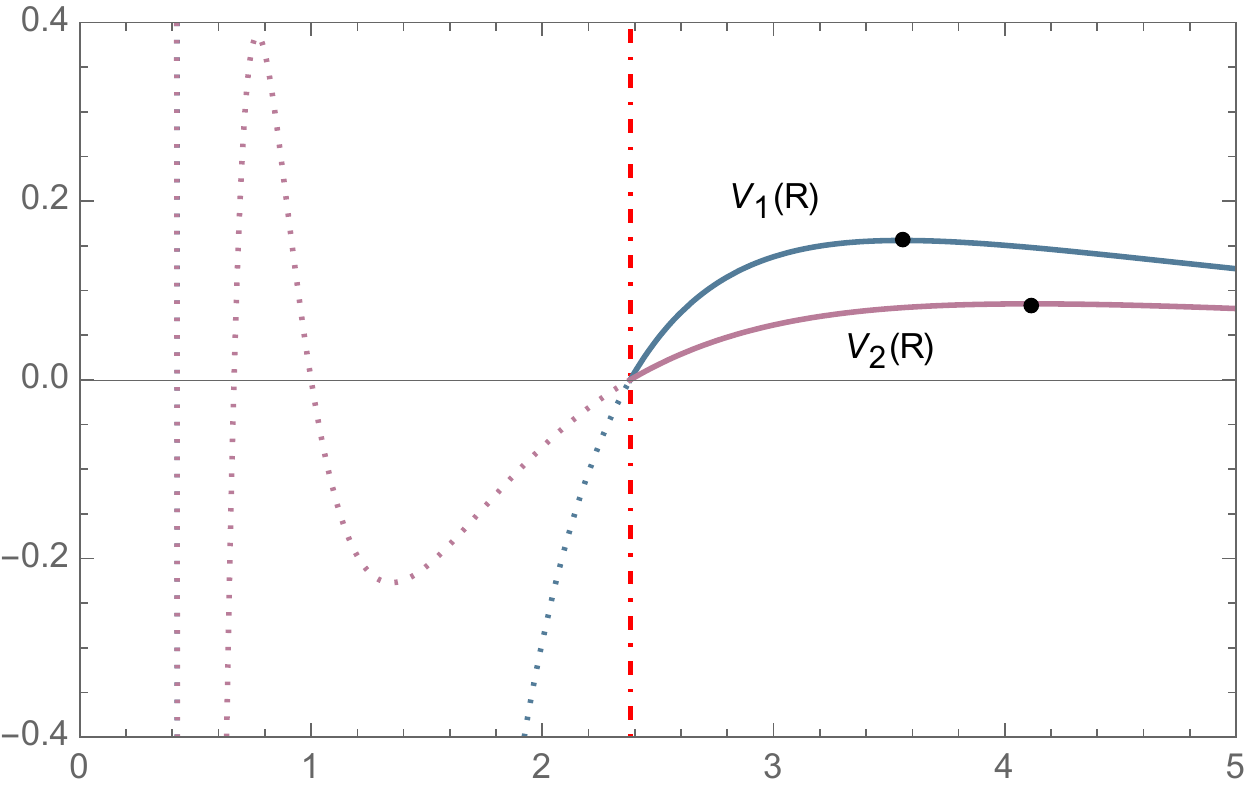}
\end{subfloat}
\caption{\label{fig8}: Quasinormal modes for \textit{undercharged} collapsed shells with {\it Left panel}: $Q=1$ and $m_0=1.02$  {\it Right panel}: $Q=1$ and $m_0=1.4$. The vertical red dotted-dashed line stands for the external horizon of the black hole.} 
\end{figure} 

\pagebreak

\section{Conclusion}
According to the WGC, there should be overcharged BHs with charge larger than their mass $|Q|>m$. Such objects are not allowed in GR as they have naked singularities, violating the cosmic censorship hypothesis. However, such objects have been constructed by extensions of GR where higher derivative terms have been included in the gravitational action. Here we examined overcharged spherical symmetric objects in GR and, in particular, we studied charged shells. The shells we considered are made of perfect fluids with tension $\sigma$ and pressure $p$ obeying $p=w\sigma$ with $|w|\leq 1$. We have found that all shells with positive pressure, $0 < w \leq 1$, whether overcharged or undercharged, are unstable and collapse to form charged black holes. For negative pressure, undercharged shells with $-1\leq w <0$ are unstable. However we found overcharged shells with $-1\leq w <0$ stabilized at finite radial distance. This result is also dictated by  the cosmic censorship, since if an overcharged shell collapses, it will form a naked singularity. 
We have also seen that for nearly overcharged shells with $0.912\lesssim m/|Q|<1$, a trapping potential is formed for the gravitational perturbations as well as for the electromagnetic ones.  This is in accordance with the fact that horizonless objects can exhibit trapped spacetime modes if they are confined in the interior of the potential barrier. 
\vskip.3in

\noindent
{\bf \large Acknowledgement:} We would like to thank C. Bachas  for extensive discussions.  A.R. is funded by the Boninchi Foundation.


\begin{appendix}

\end{appendix}



\begin{thebibliography}{99}


\bibitem{V}
C.~Vafa,
``The String landscape and the swampland,''
[arXiv:hep-th/0509212 [hep-th]].


\bibitem{P}
E.~Palti,
``The Swampland: Introduction and Review,''
Fortsch. Phys. \textbf{67}, no.6, 1900037 (2019)
[arXiv:1903.06239 [hep-th]].

\bibitem{Har}
D.~Harlow, B.~Heidenreich, M.~Reece and T.~Rudelius,
``The Weak Gravity Conjecture: A Review,''
[arXiv:2201.08380 [hep-th]].


\bibitem{AHMNV}
N.~Arkani-Hamed, L.~Motl, A.~Nicolis and C.~Vafa,
``The String landscape, black holes and gravity as the weakest force,''
JHEP \textbf{06}, 060 (2007)
[arXiv:hep-th/0601001 [hep-th]].


\bibitem{motl}
Y.~Kats, L.~Motl and M.~Padi,
``Higher-order corrections to mass-charge relation of extremal black holes,''
JHEP \textbf{12}, 068 (2007)
[arXiv:hep-th/0606100 [hep-th]].

\bibitem{HS1}
T.~Crisford, G.~T.~Horowitz and J.~E.~Santos,
``Testing the Weak Gravity - Cosmic Censorship Connection,''
Phys. Rev. D \textbf{97} (2018) no.6, 066005
[arXiv:1709.07880 [hep-th]].

\bibitem{HS2}
G.~T.~Horowitz and J.~E.~Santos,
``Further evidence for the weak gravity \textemdash{} cosmic censorship connection,''
JHEP \textbf{06} (2019), 122
[arXiv:1901.11096 [hep-th]].

\bibitem{Huben}
V.~E.~Hubeny,
`Overcharging a black hole and cosmic censorship,''
Phys. Rev. D \textbf{59} (1999), 064013
[arXiv:gr-qc/9808043 [gr-qc]].

\bibitem{vR1}
U.~Danielsson, V.~Van Hemelryck and T.~Van Riet,
``Over-extremal brane shells from string theory?,''
Class. Quant. Grav. \textbf{39}, no.23, 235001 (2022)
[arXiv:2206.04506 [hep-th]].

\bibitem{kuchar}
K. Kuchar,
``Charged shells in general relativity and their gravitational collapse,''
Czech. J. Phys., 18: 435-63 (1968).
 
\bibitem{Chase}
J.E. Chase,
``Gravitational instability and collapse of charged fluid shells,''
Nuovo Cimento B 67, 136–152 (1970).


\bibitem{LE1}
P.~LeMaitre and E.~Poisson,
``Equilibrium and stability of thin spherical shells in Newtonian and relativistic gravity,''
Am. J. Phys. \textbf{87}, no.12, 961 (2019)
[arXiv:1909.06253 [gr-qc]].


\bibitem{SF1}
M.~Sharif and F.~Javed,
``Stability of charged thin-shell and thin-shell wormholes: a comparison,''
Phys. Scripta \textbf{96}, no.5, 055003 (2021)



\bibitem{SF2}
L.~M.~Reyes, M.~Chiapparini and S.~E.~P.~Bergliaffa,
``Thermodynamical and dynamical stability of a self-gravitating charged thin shell,''
Eur. Phys. J. C \textbf{82}, no.2, 151 (2022)

\bibitem{PER1}
I.~Antoniou, D.~Kazanas, D.~Papadopoulos and L.~Perivolaropoulos,
``Stabilizing spherical energy shells with angular momentum in gravitational backgrounds,''
Int. J. Mod. Phys. D \textbf{31}, no.08, 2250064 (2022)
[arXiv:2204.14003 [gr-qc]].


\bibitem{FL1}
T.~V.~Fernandes and J.~P.~S.~Lemos,
``Electrically charged spherical matter shells in higher dimensions: Entropy, thermodynamic stability, and the black hole limit,''
Phys. Rev. D \textbf{106}, no.10, 104008 (2022)
[arXiv:2208.11127 [gr-qc]].

\bibitem{CI} 
V. De La Cruz and W. Israel, Nuovo Cimento  A 5, 744 (1967). 


\bibitem{CS}
S.~Chandrasekhar,
``The mathematical theory of black holes,'' Oxford University, New York, 1983.

\bibitem{MM}
F.~Mellor and I.~Moss,
``Stability of Black Holes in De Sitter Space,''
Phys. Rev. D \textbf{41} (1990), 403

\bibitem{Berti:2003ud}
E.~Berti and K.~D.~Kokkotas,
``Quasinormal modes of Reissner-Nordstr\"om-anti-de Sitter black holes: Scalar, electromagnetic and gravitational perturbations,''
Phys. Rev. D \textbf{67}, 064020 (2003)
[arXiv:gr-qc/0301052 [gr-qc]].

\bibitem{kk1}
K.~D.~Kokkotas,
``Pulsating relativistic stars,''
[arXiv:gr-qc/9603024 [gr-qc]].


\bibitem{cc1}
S.~Chandrasekhar and V.~Ferrari,
``On the non-radial oscillations of a star,''
Proc. Roy. Soc. Lond. A \textbf{432} (1991), 247-279

\bibitem{KK1994}
K.~.D.~Kokkotas
``Axial Modes for Relativistic Stars,''
MNRAS \textbf{268} (1994), 1015


\bibitem{MM2001}
P.~O.~Mazur and E.~Mottola,
``Gravitational Condensate Stars: An Alternative to Black Holes,''
	arXiv:gr-qc/0109035

\bibitem{MM2023}
P.~O.~Mazur and E.~Mottola,
"Gravitational Condensate Stars: An Alternative to Black Holes"
Universe \textbf{9} (2023), 2



\bibitem{TSM1999}
K.~Tominaga, M.~Saijo and K.~ Maeda
``Gravitational waves from a test particle scattered by a neutron star: Axial mode case''
Phys. Rev. D \textbf{60} (1999) no.2, 024004

\bibitem{FK2000}
V.~Ferrari and K.~D.~Kokkotas
``Scattering of particles by neutron stars: Time evolutions for axial perturbations''
Phys. Rev. D \textbf{62} (2000) no.10, 107504

\bibitem{CFP2016}
V.~Cardoso, E.~Franzin and P.~Pani
``Is the Gravitational-Wave Ringdown a Probe of the Event Horizon?''
Phys. Rev. Lett. \textbf{116} (2016) no.17, 171101

\bibitem{ADA2017}
J.~Abedi, H.~Dykaar and N.~Afshordi,
``Echoes from the abyss: Tentative evidence for Planck-scale structure at black hole horizons,''
Phys. Rev. D \textbf{96} (2017) no.8, 082004

\bibitem{kk2}
S.~H.~V\"olkel and K.~D.~Kokkotas,
``A Semi-analytic Study of Axial Perturbations of Ultra Compact Stars,''
Class. Quant. Grav. \textbf{34} (2017) no.12, 125006
[arXiv:1703.08156 [gr-qc]]; 
``Ultra Compact Stars: Reconstructing the Perturbation Potential,''
Class. Quant. Grav. \textbf{34} (2017) no.17, 175015
[arXiv:1704.07517 [gr-qc]];
``Wormhole Potentials and Throats from Quasi-Normal Modes,''
Class. Quant. Grav. \textbf{35} (2018) no.10, 105018
[arXiv:1802.08525 [gr-qc]].


\bibitem{kk3}
A.~Maselli, S.~H.~V\"olkel and K.~D.~Kokkotas,
``Parameter estimation of gravitational wave echoes from exotic compact objects,''
Phys. Rev. D \textbf{96} (2017) no.6, 064045
[arXiv:1708.02217 [gr-qc]].


\bibitem{WA2018}
Q.~Wang and N.~Afshordi
"Black hole echology: The observer's manual"
Phys. Rev. D \textbf{97} (2018) no.12, 124044


\bibitem{MTBP2019}
E.~Maggio, A.~Testa, S.~Bhagwat and P.~Pani
``Analytical model for gravitational-wave echoes from spinning remnants''
Phys. Rev. D \textbf{100} (2019) no.6, 064056






 
\end{thebibliography}
\end{document}